\documentclass[12pt] {article}
\pdfoutput=1
\usepackage[hyphens,spaces,obeyspaces]{url}
\usepackage{setspace}

\usepackage{epsfig}
\usepackage{amssymb}
\usepackage{amsmath}
\usepackage{multirow}
\usepackage{setspace}
\usepackage{datetime}
\usepackage{amsmath}
\usepackage{algorithm}
\usepackage{algpseudocode}
\usepackage{graphicx}
\usepackage{subfig} % for subfigures
\usepackage{lipsum}
\usepackage{acronym}
\usepackage{array}
\usepackage{graphicx}
\usepackage{cite}
\usepackage{graphicx}
\usepackage{epstopdf}
\usepackage{caption}
\usepackage{balance}
\usepackage{fancybox}
\usepackage{colortbl}
\usepackage{array}
\usepackage{mystyle}
\usepackage{multirow}
\usepackage[obeyspaces,hyphens,spaces]{url}
\usepackage{threeparttable}
\usepackage[english]{babel}% tables
\usepackage{longtable}
\usepackage{listings, lstautogobble}
\usepackage{booktabs}
\usepackage{threeparttable}
\usepackage[table,xcdraw]{xcolor}
\usepackage{amsthm}
\AtBeginEnvironment{thm}{\footnotesize}
\AtBeginEnvironment{defn}{\footnotesize}
\renewcommand{\baselinestretch}{1}

\providecommand{\keywords}[1]{\textbf{\textit{Keywords---}} #1}

\usepackage{stackengine}
\newsavebox\mybox
\newcommand\Includegraphics[2][]{\sbox{\mybox}{%
  \includegraphics[#1]{#2}}\abovebaseline[-.5\ht\mybox]{%
  \addstackgap{\usebox{\mybox}}}}

\newtheorem{definition}{Definition}[section]
\newtheorem{lemma}{Lemma}[section]

\newtheorem{theorem}{Theorem}[section]
\usepackage{hyperref}
\usepackage[framemethod=tikz]{mdframed}
\usepackage{enumitem}

{\begin{list}{}%
         {\setlength{\leftmargin}{#1}}%
         \item[]%
}
{\end{list}}

\newdateformat{monthyeardate}{%
  \monthname[\THEMONTH], \THEYEAR}
\begin{document}
\title{\Large{\textbf{Formal Probabilistic Analysis of Dynamic Fault Trees in HOL4} }}
\author{
Yassmeen~Elderhalli, Waqar~Ahmad, Osman~Hasan,\\ and Sofi\`ene~Tahar\vspace*{2em}\\
Department of Electrical and Computer Engineering,\\
Concordia University, Montr\'eal, QC, Canada 
\vspace*{1em}\\
\{y\_elderh,waqar,o\_hasan,tahar\}@ece.concordia.ca 
 \vspace*{3em}\\
\textbf{TECHNICAL REPORT}\\
\date{July 2018}
}
\maketitle

\newpage
\begin{abstract}

Dynamic Fault Trees (DFTs) is a widely used failure modeling technique that allows capturing the dynamic failure characteristics of systems in a very effective manner. Simulation and model checking have been traditionally used for the probabilistic analysis of DFTs. Simulation is usually based on sampling and thus its results are not guaranteed to be complete, whereas model checking employs computer arithmetic and numerical algorithms to compute the exact values of probabilities, which contain many round-off errors. Leveraging upon the expressive and sound nature of higher-order-logic (HOL) theorem proving, we propose, in this work, a formalization of DFT gates and their probabilistic behavior as well as some of their simplification properties in HOL. This formalization would allow us to conduct the probabilistic analysis of DFTs by verifying generic mathematical expressions about their behavior in HOL. In particular, we formalize the AND, OR, Priority-AND, Functional DEPendency, Hot SPare, Cold SPare and the Warm SPare gates and also verify their corresponding probabilistic expressions in HOL.  Moreover, we formally verify an important property, $Pr(X<Y)$, using the Lebesgue integral as this relationship allows us to reason about the probabilistic properties of Priority-AND gate and the \textit{Before} operator. We also formalize the notion of conditional densities in order to formally verify the probabilistic expressions of the Cold SPare and the Warm SPare gates. For illustrating the usefulness of our formalization, we use it to formally analyze the DFT of a Cardiac Assist System.
\end{abstract}

\newpage
\section{Introduction}
\label{introduction}

A Fault Tree (FT) \cite{DFT-survey} represents an effective way of graphically modeling the causes of failure in a system in the form of a rooted failure tree. A typical FT consists of a \textit{top event} representing system failure, basic failure events modeling the components failure and the FT gates, which combine the basic failure events and allow components failure to propagate to the top event. FTs are categorized as: Static FTs (SFTs) and Dynamic FTs (DFTs). SFTs capture the causes of failure in a system without considering the failure dependencies or sequences between the system components. DFTs, on the other hand, capture the failure dependencies in systems, which represent a more realistic approach to model the behavior of real-world systems.

\indent Fault Tree Analysis (FTA) can be used to examine the failure characteristics of the given system qualitatively and quantitatively. In the former analysis, the combinations and sequences of basic failure events, associated with the system components, are determined in the form of cut sets and cut sequences. While the quantitative analysis allows estimating the failure probability of the system based on component's failure probabilities among other metrics. Usually, Markov chain (MC) based analysis or algebraic approaches are used to perform DFT analysis. In the Markov chain based analysis, the DFT is first converted into its equivalent MC and then the analysis is conducted on the resulting MC. Complex systems often lead to a MC with a large number of states. The MCs of such complex systems can be analyzed using a modularization approach that divides the corresponding FT into SFT and DFT parts \cite{Modular-DFT}. The SFT part is analyzed using traditional combinatorial analysis methods, such as Binary Decision Diagrams (BDDs)\cite{DFT-survey}, while the DFT part is analyzed using MCs \cite{DFT-handbook}. This kind of modularization approach has been implemented in the Galileo tool \cite{Galileo}. In the algebraic approach, an algebra similar to the ordinary Boolean algebra is used to reduce the structure function (expression) of the \textit{top event} of the DFT \cite{Merle-thesis}. This reduced expression is then used to derive the failure probability of the given system based on the failure probabilities of DFT gates.
     
\indent Traditionally, DFTs are either analyzed by analytically deriving the system failure probability expression or using computer-based simulation tools. In the former method, firstly cut-sequences consisting of basic failure events are obtained and then the probabilistic Principle of Inclusion-Exclusion (PIE) \cite{Merle-thesis} is used to derive the probability of failure of the overall system. This kind of manual manipulation is prone to human errors and can produce erroneous results especially when dealing with large DFTs. The latter method is more extensively used due to its scalability and user friendliness. Several simulation tools are available that provide GUI editors that obtain the system FT model from the user and return the analysis results based on the assigned failure distribution to the system components at a given instant of time. However, simulation cannot be gauranteed to produce complete and accurate results due to the involvement of numerical techniques, such as Monte Carlo simulation \cite{Monte-carlo}, and pseudo random variables.  Due to the above-mentioned inaccuracies, both analytical and simulation based methods are not suitable to conduct the failure analysis of  safety-critical systems.

\indent As an accurate alternative, formal methods have been recently utilized for analyzing FTs. Probabilistic model checkers (PMC), such as STORM \cite {dehnert2017storm}, have been used to perform the quantitative analysis of DFTs \cite{ghadhab2017model}. However, due to the state-based nature of PMCs, they  cannot be used to verify generic expressions for probability of failure. In addition, their usage is only limited to exponential distributions, which in the context of reliability analysis, for example, do not consider the aging of systems components. Due to the sound nature of higher-order-logic (HOL) theorem proving, it has been successfully used to formalize basic SFT gates \cite{ahmed2015towards}, which have been in turn used to conduct the SFT-based analysis of several systems, including an air traffic management system\cite{ahmed2016formalization}. However, this formalization is only limited to SFTs. So far, there is no formalization in HOL that supports the probabilistic failure analysis of DFTs. Recently, a hybrid methodology based on both interactive theorem proving and model checking has been presented for formal analysis of DFTs \cite{Yassmeen-NFM18}. The main idea is to first conduct the algebraic based qualitative analysis of a given DFT using theorem proving and then quantitatively analyze the simplified DFT model using the STORM model checker. Since a PMC is involved in estimating the probabilities quantitatively, this methodology cannot provide generic expressions for probability and its usage is only limited to exponential distributions. Moreover, the formal definitions of DFT gates in \cite{Yassmeen-NFM18} cannot cater for conducting the probabilistic analysis using HOL theorem proving as the behavior of the DFT gates has been captured using numbers instead of random variables. \\ 

\indent In order to perform the complete probabilistic analysis of DFTs within a higher-order-logic theorem prover by verifying generic expressions of probability of failure, we propose to formalize the DFT gates in higher-order logic. Based on this formalization, we also formally verify their algebraic reduction properties. Then, using the available probability theory formalization \cite{Tarek-thesis}, we also formally verify the failure probability relationships of all commonly used DFT gates, i.e., AND, OR, Priority-AND (PAND), Functional DEPendency (FDEP), Hot SPare gate (HSP), Cold SPare gate (CSP) and Warm SPare gate (WSP). In order to verify the failure probability relationship of some of these DFT gates, we are required to formalize the $Pr (X < Y)$ describing the effect of one system component failing before the other or one after the other. This property is mainly verified by using Lebesgue Integral  properties \cite{Mhamdi-entropy, Qasim-CICM}. In addition, we formalize the notion of conditional density functions, which is necessary to formally verify the probabilistic relationships of the spare gates.  The HOL4 theorem prover \cite{HOL4} was a natural choice for this formalization as it has the required theories such as: the probability theory and the Lebesgue integral \cite{Mhamdi-entropy}. In addition, we use the existing formalization of the probabilistic PIE in HOL4   \cite{ahmed2015towards}. The above-mentioned formalizations can be utilized to conduct the DFT-based failure analysis of a variety of real-world systems within the sound core of a theorem prover. For illustration purposes, we present the formal DFT-based failure analysis of a Cardiac Assist System (CAS) \cite{hichem-rigorous2010}, which is a safety-critical DFT benchmark. We first reduce the original structure function of the system's top event using the formally verified simplification theorems. Then, we utilize the probabilistic PIE \cite{ahmed2015towards} to formally verify a generic failure probability expression of the Cardiac Assist System whereas the failure characteristics of its components are represented as generic probability distribution and density functions.

The rest of the report is structured as follows:
 Section \ref{Preliminaries} presents some preliminaries about the probability theory and the Lebesgue integral in HOL4 that will facilitate the understanding of the rest of the work. In Section \ref{Formalization_in_hol}, we present our HOL formalization of DFT gates and the corresponding simplification properties. Section \ref{probabilistic_gates} provides the verification details of the probabilistic behavior of the DFT gates.  Section \ref{Experiment results} presents the formalization of the probabilistic failure behavior of the Cardiac Assist System. Finally, we conclude the report in Section~\ref{Conclusion}.

\section{Preliminaries}
\label{Preliminaries}
In this section, we present some preliminaries that are required for the understanding of the proposed formalization. 

\subsection{Probability Theory} 

The probability theory is formalized based on the measure theory in HOL4 \cite{Mhamdi-entropy}. A measure is generally a function that designates a certain number to a set, which represents the size of this set \cite{Tarek-thesis}. It is defined as the triplet $(\sigma, \Sigma, \mu)$, where $\sigma$ represents the space, $\Sigma$ represents the measurable sets and finally $\mu$ represents the measure. A probability space is defined as a measure space, where the probability measure for the entire space is equal to 1.

Random variables are formalized as measurable functions that map events from the probability space to some other $\sigma$- algebra space $s$. Random variables are defined in HOL4 as \cite{Tarek-thesis}:

%\begin{defn}
%\label{random_variable}
%\vspace{-16pt}
%\emph{Random Variable} \\

\begin{mdframed}
	\begin{flushleft}
\begin{definition}
\label{DEF:random_variable}
\emph{}\\
{\small\textup{\texttt{$\vdash$ $\forall$X p s. random\_variable X p s $\Leftrightarrow$}\\
       \mbox {\texttt{~~~~~~~~~~~~prob\_space p $\wedge$ X $\in$ measurable (p\_space p, events p) s }}}}
\end{definition}

   \end{flushleft}
\end{mdframed}      

\vspace{1pt}\noindent where \texttt{prob\_space p} ensures that \texttt{p} is a probability space with \texttt{p\_space} as its space and events as its measurable sets.  \texttt{X $\in$ measurable (p\_space p, events p) s} defines \texttt{X} as a measurable function from the probability space \texttt{p} to space \texttt{s}.

The probability distribution of a random variable $X$ represents the probability that the random variable $X$ belongs to a set $A$. This is equivalent to finding the probability of the event $\{X \in A\}$, which can also be represented using the preimage as $X^{-1}(A)$. The probability distribution is defined in HOL4 as \cite{Tarek-thesis}:\\

\begin{mdframed}
	\begin{flushleft}
\begin{definition}
\label{DEF:distribution}
\emph{}\\
{\small\textup{\texttt{\texttt{$\vdash$ $\forall$p X. distribution p X = ($\lambda$s. prob p (PREIMAGE X s $\cap$ p\_space p)) }}}}
\end{definition}

   \end{flushleft}
\end{mdframed}      
%
%\vspace{2pt}\noindent {\texttt{$\vdash$ $\forall$ p X. distribution p X = ($\lambda$ s. prob p (PREIMAGE X s $\cap$ p\_space p)) }}\\
%%\end{defn}
%\vspace{2pt}

For a random variable that maps the probability space into another space ($s$), the push forward measure is a measure that uses the space and subsets of $s$ as its space and measurable sets and uses the distribution of the random variable as its measure part \cite{Johannes-Thesis}.  In general, the push forward measure for any measurable function $X$ from measure $M$ to measure $N$ can be expressed as :

\begin{mdframed}
	\begin{flushleft}

\begin{definition}
\label{push-forward-measure}
\emph{} \\
{\small\textup{\texttt{$\vdash$ $\forall$ M N f. distr M N f = }\\
{\texttt{~~~~(m\_space N, measurable\_sets N,}}\\
{\texttt{~~~~~~~~ $\lambda$A. measure M (PREIMAGE f A $\cap$ m\_space M))}}}}\\
\end{definition}

   \end{flushleft}
\end{mdframed}      

The cumulative distribution function (CDF) of a random variable $X$ is usually used when we are interested in finding the probability that the random variable is less than or equal to a certain value. It is formally defined as \cite{ahmed2015towards}:\\

\begin{mdframed}
	\begin{flushleft}
         \begin{definition}
\label{DEF:Cumulative_density_function}
\emph{}\\
{\small\textup{\texttt{\texttt{$\vdash$ $\forall$p X t. CDF p X t = distribution p X \{y | y $\leq$ (t:real)\}} }}}
\end{definition}
   \end{flushleft}
\end{mdframed}      

When dealing with multiple random variables, the probabilistic \textit{Principle of Inclusion and Exclusion} (PIE) provides a very interesting relationship between the probability of the union of different events. It can be expressed as:

\begin{equation}
Pr(\bigcup_{i=1}^{n} A_{i})= \sum_{t\neq\{\}, t\subseteq \{1,2,...,m\}} (-1)^{|t|+1} Pr(\bigcap_{j\in t}A_{j})
\end{equation}

It has been formally verified in HOL4 as follows \cite{ahmed2015towards}:

\begin{mdframed}
	\begin{flushleft}
\begin{theorem}
			\label{THM:PIE}
\emph{} \\{\small
				\textup{\texttt{$\vdash$ $\forall$p L1 L2.}\\
				{\texttt{~~~~prob\_space p $\wedge$ ($\forall$ x. MEM x L $\Rightarrow$ x $\in$ events p) $\Rightarrow$}\\
\mbox{\texttt{~~~~(prob p (union\_list L = sum\_set \{t | t $\subseteq$ set L $\wedge$ t $\neq$ \{\}\}  }}\\
\mbox{\texttt{~~~~($\lambda$t. -1 pow (CARD t+1) * prob p (BIGINTER t))}}}}}
		\end{theorem}
   \end{flushleft}
\end{mdframed}      	
	
\noindent where \texttt{L} is the list of events that we are interested in expressing the probability of their union.

%
%\noindent \mbox{\texttt{$\vdash$ $\forall$ p L1 L2. prob\_space p $\wedge$ ($\forall$ x. MEM x L $\Rightarrow$ x $\in$ events p) $\Rightarrow$}}\\
%\mbox{\texttt{~~~~(prob p (union\_list L = sum\_set \{t | t $\subseteq$ set L $\wedge$ t $\neq$ \{\}\}  }}\\
%\mbox{\texttt{~~~~($\lambda$t. -1 pow (CARD t+1) * prob p (BIGINTER t))}}
\vspace{1pt}

In order to be able to handle multiple random variables, a pair measure (often called binary prodcut measure) is required to be able to model joint distribution measures. This pair measure can be used also in a nested way to model the joint distribution measure of multiple random variables. The pair measure is defined as the product of two measures. It was initially formalized in Isabelle/HOL \cite{Johannes-Thesis} and was then ported to HOL4 \cite{Qasim-thesis}. 
The space and the measurable sets of this pair measure are generated using the Cartesian product of the spaces and the measurable sets of the participating measures, while the measure part is defined using the Lebesgue integral.
 
The Lebesgue-Borel measure is required to integrate over the real line. In particular, we need the Lebesgue-Borel measure in this work to integrate the density functions of the random variables over the real line.  The Lebesgue-Borel measure is a measure defined over the real line, which uses the real line as its space and the borel sets as its measurable sets. The Lebesgue-Borel measure is defined in HOL4 as \texttt{lborel}, which uses the real borel sigma algebra generated by the open sets of the real line as well as the Lebesgue measure \cite{Qasim-CICM}.

\subsection{Lebesgue Integral}

The Lebesgue integral is defined in HOL4 using positive simple functions, which are measurable functions defined as a linear combinations of indicator functions of measurable sets representing a partition of the space $X$ \cite {Mhamdi-entropy}. A positive simple function, $g$, can be represented using the triplet $(s,\ a, \ x)$ as \cite {Mhamdi-entropy}:

\begin{equation}
\forall  t \in X,\ g(t) = \sum _{i \in s} x_{i} \textbf{1} _{a_{i}} (t), \ \ \ \ x_{i} \geq 0
\end{equation}

\noindent where $\textbf{1}_{a_{i}}$ is the indicator function of measurable set $a_{i}$ and is defined as \cite{Mhamdi-entropy}:\\
%\begin{defn}
%\label{indicator_fn}
%\vspace{-16pt}
%\emph{indicator function} \\

\begin{mdframed}
	\begin{flushleft}
      \begin{definition}
\label{DEF:indicator_fn}
\emph{}\\
{\small\textup{\texttt{\texttt{$\vdash$ $\forall$A. indicator\_fn A = ($\lambda$ x. if x $\in$ A then 1 else 0)}} }}
\end{definition}
   \end{flushleft}
\end{mdframed}      

The Lebesgue integral is first defined for positive simple functions and then extended for positive functions for measure $\mu$ as \cite{Tarek-ITP2010}:\\ 
%\begin{equation}
%\int_{X}g\ d \mu = \sigma_{i \in s} x_{i} \mu (a_{i})
%\end{equation}
%
%\noindent where measure $\mu$ is defined as:\\
%This integral is extended to positive functions as\cite {Mhamde-entopy}:
%
\begin{equation}
\int_{X} f d\mu = sup \{\int_{X} g\ d\mu\ |\ g\ \leq\ f\ and\ g\ positive\ simple\ function\}
\end{equation}

It is usually required that the probability of an event for a continuous random variable to be expressed using the integration of the random variable's distribution. This is verified in HOL4 as\cite {Tarek-thesis}:
 
%\begin{thm}
%\label{distribution-integrarion}
%\vspace{-16pt}

\begin{mdframed}
	\begin{flushleft}
		\begin{theorem}
			\label{THM:distribution_integral}
\emph{} \\{\small
				\textup{\texttt{$\vdash$ $\forall$X p s A.}\\
				{\texttt{~~~~random\_variable X p s $\wedge$ A $\in$ subsets s $\Rightarrow$ } \\
{\texttt{~~~~(distribution p X A = }}\\
{\texttt{~~~~~integral (space s, subsets s, distribution p X)(indicator\_fn A))}}}}}
		\end{theorem}
   \end{flushleft}
\end{mdframed}      
\section{Formalization of Dynamic Fault Trees in HOL}
\label{Formalization_in_hol}
Our previous formalization of DFT gates and operators was based on the algebraic approach \cite{Merle-thesis}, where the DFT events are treated based on their time of occurrence (failure of corresponding components)\cite{Yassmeen-NFM18}. However, these definitions cannot cater for the probabilistic analysis of system failures, which is the scope of the current work. Therefore, we provide an improved formalization of DFT gates and operators using functions of time that can be represented as random variables when carrying out the formal probabilistic analysis of the given DFT.  \\
\subsection{Identity Elements and Temporal Operators}

Similar to ordinary Boolean algebra, the DFT algebraic approach defines identity elements that are important in the simplification process of the DFT \cite{Merle-thesis}. The DFT identity elements are: the \emph{ALWAYS} event representing an event that always occurs (fails) from time 0 and the \emph{NEVER} event, which describes an event that never occurs (fails). The formal definitions of these elements are shown in Table~ \ref{table:element-operator}, where \texttt{extreal} is the extended real numbers datatype in HOL4 and  \texttt{PosInf} represents $+\infty$ in HOL4. We define the events as lambda abstracted functions so that they can be later treated as random variables.

\begin{table}[b]
\caption{Definitions of Identity Elements and Temporal Operators}
\vspace{5pt}
\scriptsize
\centering
\label{table:element-operator}
\begin{tabular}{|p{2.4cm}|p{3.9cm} | p{6.25cm}|}
\hline
Element/Operator & Mathematical Expression  &  Formalization \\ \hline  \hline
{\scriptsize{\texttt{Always element}}} & 
 $\!\begin{aligned}[b]
		{\scriptstyle d(ALWAYS)\ =\ 0}
	\end{aligned}$&$\!\begin{aligned}[c]
	& \scriptsize{\texttt{$\vdash$ ALWAYS = ($\lambda$s. (0:extreal))  }}\end{aligned}$  \\ \hline
{\scriptsize{\texttt{Never element}}} &	
$\!\begin{aligned}[b]
	{\scriptstyle d(NEVER)\ =\ \texttt{+$\infty$}}
	\end{aligned}$& $\!\begin{aligned}[c]
	& \scriptsize{\texttt{$\vdash$ NEVER = ($\lambda$s. PosInf)  }}\end{aligned}$  \\ \hline
{\scriptsize{\texttt{Before}}}&
$\!\begin{aligned}[b]
	{{\scriptstyle d(X  \lhd Y)= }{\tiny 
	\begin{cases}  d(X), &d(X) < d(Y)\\ +\infty, &d(X)\geq d(Y)
\end{cases}} 
}
	\end{aligned}$& $\!\begin{aligned}[c]
	& \scriptsize{\texttt{$\vdash$ $\forall$X Y. 
		D\_BEFORE X Y =}}\\[-1\jot]
		&\scriptsize{\texttt{($\lambda$s. if X s  < Y s then X s else PosInf)
}}\end{aligned}$   
 \\ \hline
{\scriptsize{\texttt{Simultaneous}}}& $\!\begin{aligned}[b]
	{{\scriptstyle d(X  \Delta Y)= }{\tiny
	\begin{cases}  d(X), &d(X) = d(Y)\\ +\infty, &d(X)\neq d(Y)
\end{cases}} 
}
	\end{aligned}$ & $\!\begin{aligned}[c]
	& \scriptsize{\texttt{$\vdash$ $\forall$X Y. 
		D\_SIMULT X Y =}}\\[-1\jot]
		&\scriptsize{\texttt{($\lambda$s. if X s  = Y s then X s else PosInf)
}}\end{aligned}$    \\ \hline
{\scriptsize{\texttt{Inclusive Before}}}& $\!\begin{aligned}[b]
	{{\scriptstyle d(X  \unlhd Y)=}{\tiny  
	\begin{cases}  d(X), &d(X) \leq d(Y)\\ +\infty, &d(X) > d(Y)
\end{cases}} 
}
	\end{aligned}$ & $\!\begin{aligned}[c]
	& \scriptsize{\texttt{$\vdash$ $\forall$ X Y. 
		D\_INCLUSIVE\_BEFORE X Y =}}\\ 		&\scriptsize{\texttt{($\lambda$s. if X s  $\leq$ Y s then X s else PosInf)
}}\end{aligned}$    \\ \hline
\end{tabular}

\end{table}
%%
%%$\!\begin{aligned}[b]
%		{\scriptstyle d(ALWAYS)\ =\ 0}
%	\end{aligned}$&$\!\begin{aligned}[c]
%	& \scriptsize{\texttt{$\vdash$ ALWAYS = ($\lambda$s. (0:extreal))  }}\end{aligned}$  \\ \hline
%
%\begin{equation}
%d(A  \lhd B) =  \begin{cases}d(A), &d(A) < d(B)\\ +\infty, &d(A)\geq d(B) 
%\end{cases} 
%\end{equation}
%%
%%{p{.5\textwidth}c}
%\begin{defn}
%\label{ALWAYS_def}
%%\vspace{-16pt}
%\emph{ALWAYS element} \\
%\vspace{1pt} {\texttt{$\vdash$  ALWAYS = ($\lambda$s. (0:extreal)) }}\\
%\end{defn}
%\begin{defn}
%%\texttt{\bf{Definition 3: }}
%\label{NEVER_def}
%\vspace{-14pt}
%\emph{NEVER element} \\
%\vspace{1pt} {\texttt{$\vdash$ NEVER = ($\lambda$s. PosInf)}}\\
%\end{defn}
Temporal operators are also required to model the DFT gates in the algebraic approach \cite{Merle-thesis}. These operators: are  \emph{Before}~(\(\lhd\)), \emph{Simultaneous}~(\(\Delta\)) and \emph{Inclusive Before}~(\(\unlhd\)). Each one of these operators accepts two inputs, which can be subtrees or basic events. The output event of the operator occurs according to a certain sequence of occurrence for the input events, i.e., the time of occurrence of the first (left) input is less than, equal to or less than or equal to the occurrence time of the second input (right) for the \emph{Before}, the \emph{Simultaneous} and the \emph{Inclusive Before}  operators, respectively. The time of occurrence of the output event of all operators is equal to the time of occurrence of the first input event (left). The mathematical expressions of these operators as well as their corresponding HOL formalizations are shown in Table~\ref{table:element-operator}, where \texttt{X} and \texttt{Y} represent the time of occurrence of events X and Y, respectively. 

It is worth mentioning that if the inputs of the \emph{Simultaneous} operator are basic events with continuous failure distributions, then the output of this operator can never fail \cite{Merle-thesis}. This can be expressed for basic events $X$ and $Y$ as:
 \begin{equation}
 \label{simult_never}
d(X  \Delta Y) =  NEVER
\end{equation}
%\vspace{-20pt}
\subsection{Formalization of FT Gates and Simplification Theorems}

\indent Our formalization of all FT gates; static and dynamic, and their mathematical expressions are presented in Table \ref{table:gates}.
%\vspace{-5pt}
\begin{table}[!b]
\centering
\small
%\vspace{-20pt}
\caption{DFT Gates}
\vspace{5pt}
\label{table:gates}
\begin{tabular}{|p{1.6cm}|p{5.0cm} | p{6.1cm}|}
\hline
Gate & Mathematical Expression & Formalization\\ \hline \hline
{\Includegraphics[scale=0.3]{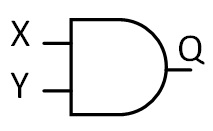}}                    & \multirow{2}{*}{$\!\begin{aligned}[b]
	{{\scriptsize d(X \cdot Y) = max (d(X),d(Y))
} 
}
	\end{aligned}$  }  & \multirow{2}{*}{$\!\begin{aligned}[l]
	&\scriptsize{\texttt{$\vdash$ $\forall$X Y. D\_AND X Y = ($\lambda$s. max (X s)(Y s))
}}\end{aligned}$}  \\
\scriptsize AND                        &                            &                           \\ \hline
{\Includegraphics[scale=0.3]{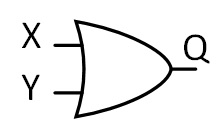}} & \multirow{2}{*}{$\!\begin{aligned}[b]
	{{\scriptsize d(X + Y) = min (d(X),d(Y))
} 
}
	\end{aligned}$}   & \multirow{2}{*}{$\!\begin{aligned}[l]
	& \scriptsize{\texttt{$\vdash$ $\forall$X Y. D\_OR X Y = ($\lambda$s. min (X s)(Y s))
}}\end{aligned}$}   \\
\scriptsize OR                         &                            &                           \\ \hline
{\Includegraphics[scale=0.3]{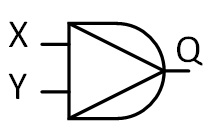}}                 & \multirow{2}{*}
{$\!\begin{aligned}[b]
	{{\scriptsize d(Q_{PAND})= }{\scriptsize 
	\begin{cases}  d(Y), &d(X) \leq d(Y)\\ +\infty, &d(X) > d(Y)
\end{cases}} 
}
	\end{aligned}$} & \multirow{2}{*}{$\!\begin{aligned}[t]
	& \scriptsize{\texttt{$\vdash$ $\forall$X Y. PAND X Y =}}\\[-2\jot]
	&\scriptsize{\texttt{($\lambda$s. if X s $\leq$ Y s then Y s else PosInf)
}}\end{aligned}$} \\
\scriptsize PAND                       &                            &                           \\ \hline
{\Includegraphics[scale=0.3]{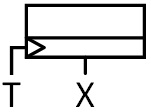}}                 & \multirow{2}{*}{$\!\begin{aligned}[b]
	{{\scriptsize d(X_{T}) = min (d(X),d(T))
} 
}
	\end{aligned}$} & \multirow{2}{*}{$\!\begin{aligned}[l]
	& \scriptsize{\texttt{$\vdash$ $\forall$X T. 
		FDEP X T = ($\lambda$s. min (X s)(T s))
}}\end{aligned}$} \\
\scriptsize FDEP                       &                            &                           \\ \hline
\multirow{2}{*}{{\Includegraphics[scale=0.3]{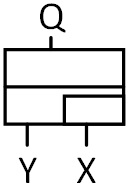}}} & $\!\begin{aligned}[b]
	{{\scriptsize d(Q_{CSP})= }{\scriptsize 
	\begin{cases}  d(X), &d(Y) < d(X)\\ +\infty, &d(Y) \geq d(X)
\end{cases}} 
}
	\end{aligned}$                   & $\!\begin{aligned}[c]
	& \scriptsize{\texttt{$\vdash$ $\forall$X Y. 
		CSP Y X =}}\\[-2\jot]
		&\scriptsize{\texttt{ ($\lambda$s. if Y s  < X s then X s else PosInf)
}}\end{aligned}$                 \\ \cline{2-3} 
                           &  $\!\begin{aligned}[b]
	{{\scriptsize d(Q_{HSP}) = max (d(Y),d(X))
} 
}
	\end{aligned}$                   & $\!\begin{aligned}[c]
	& \scriptsize{\texttt{$\vdash$ $\forall$X Y. 
		HSP Y X = ($\lambda$s. max (Y s)(X s))
}}\end{aligned}$                \\ \cline{2-3} 
\scriptsize \vspace{5pt}Spare                      & $\!\begin{aligned}[t]
	{{\scriptsize d(Q_{WSP}) =  d(Y\cdot(X_{d} \lhd Y)+}}\\{{\scriptsize 
	 X_{a}\cdot(Y \lhd X_{a})+}}\\
	 {{\scriptsize Y \Delta X_{a} + Y \Delta X_{d}    
} 
}
	\end{aligned}$& 
 {$\!\begin{aligned}[t]
	& \scriptsize{\texttt{$\vdash$ $\forall$Y X\_a X\_d. WSP Y X\_a X\_d = }}\\[-2\jot]
	&\scriptsize{\texttt{    D\_OR(D\_OR(D\_OR (D\_AND Y (D\_BEFORE X\_d Y))}}\\[-2\jot]	
	&\scriptsize{\texttt{(D\_AND X\_a (D\_BEFORE Y X\_a)))}}\\[-2\jot]
	&\scriptsize{\texttt{(D\_SIMULT Y X\_a))(D\_SIMULT Y X\_d)
		}}\end{aligned}$}                                 \\ \hline
		\multirow{2}{*}{\Includegraphics[scale=0.25]{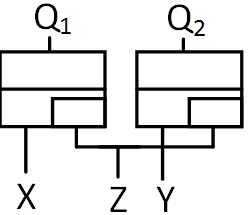}} & \multirow{3}{*}{$\!\begin{aligned}[t]
	{{\scriptsize d(Q_{1}) = d(X\cdot (Z_{d} \lhd X) +}}\\ 
{{\scriptsize Z_{a}\cdot (X \lhd Z_{a}) +}}\\{{\scriptsize X\cdot (Y \lhd X))
} 
}
	\end{aligned}$} & \multirow{3}{*}{{$\!\begin{aligned}[t]
	& \scriptsize{\texttt{$\vdash$ $\forall$X Y Z\_a Z\_d.}}\\[-2\jot]
	&\scriptsize{\texttt{   shared\_spare X Y Z\_a Z\_d =  }}\\[-2\jot]
	&\scriptsize{\texttt{   D\_OR (D\_OR (D\_AND X (D\_BEFORE Z\_d X))}}\\[-2\jot]
	&\scriptsize{\texttt{   (D\_AND Z\_a (D\_BEFORE X Z\_a)))}}\\[-2\jot]	
	&\scriptsize{\texttt{   (D\_AND  X (D\_BEFORE Y X)))
		}}\end{aligned}$}                                 } \\
                                   &                                     &                                    \\
\scriptsize \vspace{7pt} \vspace{5pt}Shared Spare            &                                     &                                    \\ \hline

\end{tabular}
\end{table}
%%
%%Q_{1} = A.(C_{d} \lhd A) + C_{a}.(A \lhd C_{a}) + A . (B \lhd A)
%\vspace{1pt}\noindent{\texttt{$\vdash$ $\forall$A B C\_a C\_d. \\}}\mbox{\texttt{~~~~shared\_spare A B C\_a C\_d =}}\\\mbox{\texttt{~~~~D\_OR (D\_OR (D\_AND A (D\_BEFORE C\_d A))(D\_AND C\_a (D\_BEFORE A C\_a)))}}\\\mbox{\texttt{~~~~~~(D\_AND  A (D\_BEFORE B A))) }}
\subsubsection{AND and OR Gates}
\noindent The AND ($\cdot$) and OR (+) gates can be modeled based on the time of occurrence of their output events. For the AND gate, the output occurs when both of its input events occur and the time of occurrence of the output is modeled as the maximum time of occurrence of both input events. For the OR gate, the output occurs once one of its input events occurs. Therefore, we formalize it as the minimum time of occurrence of the inputs. In Table \ref{table:gates}, \texttt{max} and \texttt{min} are the HOL4 functions that represent the maximum and the minimum functions, respectively.

\subsubsection{Priority AND Gate (PAND)}

\indent The PAND gate, shown in Table \ref{table:gates}, captures the sequence of occurrence (failure) of its inputs. The output event of this gate occurs if all input events occur in a certain sequence (conventionally from left to right). The behavior of the PAND can also be represented  using the temporal operators as:

\begin{equation}
\label{eq:PAND1}
Q =  Y\cdot (X \unlhd Y)
\end{equation}
%\begin{defn}
%\label{PAND}
%\emph{PAND gate} \\
%\vspace{1pt}\noindent {\texttt{$\vdash$ $\forall$A B.
%PAND A B = ($\lambda$s. if A s \(\leq\) B s  then B s else PosInf)
%}}
%\end{defn}\vspace{-2pt}
%
\noindent We verify the above relationship in HOL4 as follows: 
%\vspace{-2pt}

\begin{mdframed}
	\begin{flushleft}
		\begin{theorem}
			\label{THM:PAND2}
             {\small
	          \textup{\texttt{$\vdash$ $\forall$X Y. PAND X Y = D\_AND Y (D\_INCLUSIVE\_BEFORE X Y)
}}}
		\end{theorem}
	\end{flushleft}
\end{mdframed}

\noindent This result ascertains that the behavior of PAND gate is correctly captured in our formal definition.

\subsubsection{Functional DEPdency Gate (FDEP)}

\indent The FDEP is used to model the dependencies in the failure behavior between the system components. In other words, it is used when the failure of one component triggers the failure of another. For the FDEP gate, shown in Table \ref{table:gates}, event $X$ can occur if it is triggered by the failure of $T$ or if it occurs by itself. As a result, the occurrence time of $X_{T}$ (triggered $X$) equals the minimum time of occurrence of $T$ and $X$. From the FDEP definition, we can notice that its behavior is equivalent to the behavior of the OR gate. We verify this in HOL4 as follows: 

\begin{mdframed}
	\begin{flushleft}
		\begin{theorem}
			\label{THM:FDEP2}
          {\small
	          \textup{\texttt{$\vdash$ $\forall$X T. FDEP X T = D\_OR X T
	}}}
		\end{theorem}
	\end{flushleft}
\end{mdframed}
\subsubsection{Spare Gates}
Modeling spare parts in real systems is necessary when analyzing the probability of failure of the overall system, as these spares are used to replace the main parts after their failure. The main part $Y$ of the spare gate, shown in Table \ref{table:gates}, is replaced by the spare part $X$ after a failure occurs. The spare gate has three variants depending on the type of the spare: 
\begin{itemize}
\item\textbf{{\textit{Cold SPare Gate (CSP)}}}: The spare part can only fail while it is active.
\item \textbf{{\textit{Hot SPare Gate (HSP)}}}: The spare part can fail in both the active and the dormant states with the same probability.
\item\textbf{{\textit{Warm SPare Gate (WSP)}}}: The spare part can fail in both the dormant and active states with different probabilities. 
\end{itemize}
While manipulating the structure function of the DFT, it is required to distinguish between the two states of the spare part, i.e., the active state and the dormant state, therefore a different variable is assigned to each state. For example, for the spare gate in Table~\ref{table:gates}, variable $X$ is assigned $X_{d}$ and $X_{a}$ for the dormant and active states, respectively. 

It can be noticed from the definition of the $WSP$ gate that the  output of the spare occurs in two cases; if the spare fails in its dormant state, then the main part fails or the main part fails then the spare is activated and then it fails in its active state. The last two terms in the WSP definition cover the possibility that the spare and the main part fail at the same time. This can happen if the main part and the spare are functionally dependent on the same trigger.  For the $CSP$ gate, the output occurs if the main part fails then the spare is activated and then the spare fails while it is active. 
Since the spare part of the $HSP$ has the same failure distribution in both of its states, the output of the $HSP$ occurs when both inputs (main and spare) fail. We formally verify in HOL4 that the $HSP$ is equivalent to an AND gate as:

\begin{mdframed}
	\begin{flushleft}
		\begin{theorem}
			\label{THM:HSP_AND}
             {\small
	          \textup{\texttt{$\vdash$ $\forall$X Y. HSP Y X = D\_AND Y X
	}}}
		\end{theorem}
	\end{flushleft}
\end{mdframed}

%\vspace{-8pt}
In some real-world applications, a spare part can replace one of two main parts. This case is represented using shared spare gates as shown in Table \ref{table:gates} \cite{Yassmeen-NFM18}. The expression of the output $Q_{1}$ of the first gate is listed in Table \ref{table:gates}. This expression implies that the output $Q_{1}$ of this gate occurs in three different situations: \textit{(i)} if the main part $X$ fails, then the spare fails while it is active ($Z_{a}$), \textit{(ii)} if the spare part fails in its dormant state $Z_{d}$, then the main part fails, or \textit{(iii)} if the second main part (of the other gate) $Y$ fails before $X$, and thus the spare is not available to replace $X$ when it fails. We use the DFT operators to model the behavior of this gate, as shown in Table~ \ref{table:gates}.\\
%\subsection{Formal Verification of the Simplification Theorems}
%\label{Simplification}
\indent In the DFT algebraic approach, many simplification theorems exist and are used to reduce the structure function of the top event. In \cite{Yassmeen-NFM18}, we verified over 80 simplification theorems. However, these theorems were based on our old definitions of the DFT gates and operators that cannot cater for probabilistic analysis. We verify all these theorems for the new definitions, presented in this work, and the details can be accessed from \cite{ICFEM-code}. These simplification theorems range from simple ones, such as commutativity of the AND, OR and Simultaneous operator, to more complex ones that include combinations of all the operators. Table \ref{table:all_theorems} includes some of these verified properties.
 
%, like $(X\lhd Y)+(X \Delta Y)+(X.(Y \lhd X))\ =\ X $.
%\vspace{-3pt}
\begin{table}[]
%\vspace{-20pt}
\centering 
%\setstretch{0.7}
\caption{Examples of Formally Verified Simplification Theorems \label{table:all_theorems}}
\vspace{5pt}
\begin{tabular}{ | p{4.6cm} | p{8cm} |}
\hline
  DFT Algebra Theorems & HOL Theorems\\ \hline

%
%\begin{longtable}{ | p{7.4cm} | p{4.6cm} |}
%
%\caption{Some Verified Simplification Theorems \label{table:all_theorems}}\\
%% 
%% %\hline
%%%\multicolumn{2}{| c |}{Begin of Table}\\
%% \hline
%% HOL Formalization & Corresponding Theorem\\
%% \hline
%% \endfirsthead
%% 
%% %\hline
%% %\multicolumn{2}{|c|}{Continuation of Table \ref{long}}\\
%% \hline
%% HOL Formalization &  Corresponding Theorem\\
%% \hline
%% \endhead
%% 
%% \hline
%% \endfoot
% 
%% \hline
% %\multicolumn{2}{| c |}{End of Table}\\
% %\hline\hline
% %\endlastfoot
% 
\hline
	 $\!\begin{aligned}[b]
	{\scriptstyle X+Y = Y+X}
	\end{aligned}$&$\!\begin{aligned}[c]
	& \footnotesize{\texttt{$\vdash$ $\forall$X Y. D\_OR X Y = D\_OR Y X  }}\end{aligned}$ \\	\hline
%%	
%% 	 $\!\begin{aligned}[b]
%%	  {\scriptstyle A.B = B.A}
%%	\end{aligned}$&$\!\begin{aligned}[c]
%%	&  \footnotesize{\texttt{$\vdash$ $\forall$A B. D\_AND A B = D\_AND B A }}\end{aligned}$ \\	\hline
%%%	
%%	$\!\begin{aligned}[b]
%%	      {\scriptstyle A+A = A}
%%	\end{aligned}$&
%%  	$\!\begin{aligned}[c]
%%	& \footnotesize{\texttt{$\vdash$ $\forall$A. D\_OR A A = A }}\end{aligned}$ \\	\hline
%%	
%%	
	$\!\begin{aligned}[b]
	       {\scriptstyle X.NEVER = NEVER}
	\end{aligned}$&
 $\!\begin{aligned}[c]
	& \footnotesize{\texttt{$\vdash$ $\forall$X. D\_AND X NEVER = NEVER}}\end{aligned}$ \\	\hline
%
%%
%%$\!\begin{aligned}[b]
%% {\scriptstyle A \Delta B = B \Delta A}
%%	\end{aligned}$&
%%$\!\begin{aligned}[c]
%%	& \footnotesize{\texttt{$\vdash$ $\forall$ A B. D\_SIMULT A B = D\_SIMULT B A}}\end{aligned}$  \\	\hline
%
%
%$\!\begin{aligned}[b]
%	         {\scriptstyle (A \lhd B).(B \lhd A) = NEVER}
%	\end{aligned}$ &
%   $\!\begin{aligned}[c]
%	& \footnotesize{\texttt{$\vdash$ $\forall$ A B. D\_AND (D\_BEFORE A B)}}\\[-2\jot]
%		&\footnotesize{\texttt{(D\_BEFORE B A) = NEVER}}\end{aligned}$ \\	\hline
%%	
%%
 $\!\begin{aligned}[b]
{\scriptstyle X \lhd (Y+Z) = (X \lhd Y).( X \lhd Z)}
	\end{aligned}$&
 $\!\begin{aligned}[c]
	& \footnotesize{\texttt{$\vdash$ $\forall$ X Y Z. D\_BEFORE X (D\_OR Y Z) =}}\\[-2\jot]
	&\footnotesize{\texttt{D\_AND (D\_BEFORE X Y)(D\_BEFORE X Z)}}\end{aligned}$ \\	\hline
%%%	
%%	
%%%3.34
%%$\!\begin{aligned}[c]
%%	& \footnotesize{\texttt{$\vdash$ $\forall$ A B C. D\_AND (D\_AND (D\_BEFORE A B)}}\\
%%	&\footnotesize{\texttt{(D\_BEFORE B C))(D\_BEFORE A C) =}}\\
%%	&\footnotesize{\texttt{D\_AND (D\_BEFORE A B)(D\_BEFORE B C)}}\end{aligned}$& $\!\begin{aligned}[b]
%%{\scriptstyle (A \lhd B).(B \lhd C).(A \lhd C) = }\\ {\scriptstyle(A \lhd B).(B \lhd C)}
%%	\end{aligned}$ \\	\hline
%
%
%%$\!\begin{aligned}[c]
%%	& \footnotesize{\texttt{$\vdash$ $\forall$ A B C. D\_SIMULT A (D\_OR B C) =}}\\
%%	&\footnotesize{\texttt{ D\_OR (D\_AND (D\_SIMULT A B)}}\\
%%	&\footnotesize{\texttt{(D\_INCLUSIVE\_BEFORE B C))(D\_AND}}\\
%%	&\footnotesize{\texttt{(D\_SIMULT A C)(D\_INCLUSIVE\_BEFORE C B))}}\end{aligned}$& $\!\begin{aligned}[b]
%%{\scriptstyle A \Delta (B + C) = (A \Delta B).(B \unlhd C)+}\\ {\scriptstyle(A \Delta C).(C \unlhd B)}
%%	\end{aligned}$ \\	\hline
%
%%
%% $\!\begin{aligned}[b]
%% {\scriptstyle (A \unlhd B).(B \unlhd A) = A \Delta B}
%%	\end{aligned}$&
%%$\!\begin{aligned}[c]
%%	& \footnotesize{\texttt{$\vdash$ $\forall$ A B. D\_AND (D\_INCLUSIVE\_BEFORE A B)}}\\[-2\jot]
%%	&\footnotesize{\texttt{(D\_INCLUSIVE\_BEFORE B A) = D\_SIMULT A B}}\end{aligned}$ \\	\hline
%%	
%%
%%%3.52
 $\!\begin{aligned}[b]
 {\scriptstyle X \unlhd (Y+Z)=(X \unlhd Y).(X \unlhd Z)}
	\end{aligned}$&$\!\begin{aligned}[c]
	& \footnotesize{\texttt{$\vdash$ $\forall$ X Y Z. D\_INCLUSIVE\_BEFORE X (D\_OR Y Z) =}}\\
	&\footnotesize{\texttt{D\_AND (D\_INCLUSIVE\_BEFORE X Y)}}\\
	&\footnotesize{\texttt{(D\_INCLUSIVE\_BEFORE X Z)}}\end{aligned}$ \\	\hline
%%
%%
%%
%%	
%%%3.68
%%
	$\!\begin{aligned}[b]
{\scriptstyle (X\unlhd Y)+(X \Delta Y) = X \unlhd Y}
	\end{aligned}$&
	$\!\begin{aligned}[c]
	& \footnotesize{\texttt{$\vdash$ $\forall$X Y. D\_OR (D\_INCLUSIVE\_BEFORE X Y)}}\\[-2\jot]
	&\footnotesize{\texttt {(D\_SIMULT X Y) = D\_INCLUSIVE\_BEFORE X Y}}\end{aligned}$\\
	\hline		
%%	
%%
%%	
%%3.76
%
%	$\!\begin{aligned}[b]
%	{\scriptstyle (A\lhd B)+(A \Delta B)+(A.(B \lhd A))= A}
%	\end{aligned}$&
%	$\!\begin{aligned}[c]
%	& \footnotesize{\texttt{$\vdash$ $\forall$A B. D\_OR (D\_OR (D\_BEFORE A B)}}\\[-2\jot]
%	&\footnotesize{\texttt {(D\_SIMULT A B))(D\_AND A D\_BEFORE B A)) = A}}\end{aligned}$\\
%	\hline		
		\end{tabular}
\end{table}
%\end{longtable}
%%\vspace{-10pt}
\section{Formal Verification of DFT Probabilistic Behavior}
\label{probabilistic_gates}

In order to formally verify the probability of failure of the top event of a DFT, it is required to formally model and verify the probability of failure expression for each DFT gate. We assume that the basic events of the DFT are independent. However, in some cases these events can be dependent; in particular in the case of CSP and WSP, where the failure of the main part affects the operation and failure of the spare part. We handle this by first introducing the probabilistic behavior of the gates for independent events, then we  present the probabilistic behavior of the $WSP$ and the $CSP$ gates, which deal with dependent events.

\subsection{Probabilistic Behavior of Gates with Independent Events}
\label{Pr_independent_events}

Assuming that we are interested in finding the probability of failure until time t, the following four expressions can be used to express the probability of any DFT gate with independent basic events\cite{Merle-thesis}:\\
\begin{subequations}
\label{eq:probability_gates}
\begin{align}
Pr\{X\cdot Y\}(t) &= F_{X}(t) \times F_{Y}(t) \label{eq:pr_AND}\\
Pr\{X + Y\}(t) &= F_{X}(t) + F_{Y}(t) - F_{X}(t) \times F_{Y}(t)\label{eq:pr_OR}\\
Pr\{Y\cdot (X \lhd Y)\}(t) &=\int_{0}^{t}f_{Y}(y)\ F_{X}(y)\ dy \label{eq:pr_after}\\
Pr\{X\lhd Y\}(t) &= \int_{0}^{t} f_{X}(x)(1-F_{Y}(x))\ dx \label{eq:pr_before}
\end{align}
\end{subequations}
\noindent where $F_{X}$ and $F_{Y}$ represent the CDFs of the random variables $X$ and $Y$, respectively, and $f_{X}$ and $f_{Y}$ represent their corresponding PDFs.\\

Equation (\ref{eq:pr_AND}) represents the probability of the AND and HSP gates, which results from the probability of intersection of two independent events. Equation~(\ref{eq:pr_OR}) describes the probability of the OR and FDEP gates, which corresponds to the probability of union of two independent events. Equation (\ref{eq:pr_after}) represents the probability of having two basic events occurring in sequence one \textit{after} the other until time $t$, i.e., $Pr(X < Y)$  until time $t$ or $Pr(X < Y \wedge Y \leq t)$, which is the failure probability of the PAND for basic events. Finally, the probability of the \emph{Before} operator is represented by Equation~(\ref{eq:pr_before}), which is the probability of having event $X$ occurring \textit{before} event $Y$ until time $t$, i.e., $Pr(X < Y \wedge X \leq t)$. The difference between the last two events (\textit{before} and \textit{after}) is that in the \textit{before} event, we are just interested in finding the probability of failure of  $X$ until time $t$ with the condition that $X$ fails before $Y$. So, it is not necessary that $Y$ fails. While in the \textit{after} event, we find the probability of failure of $Y$ until time $t$ with the condition that $Y$ fails after $X$. So, it is required that both $X$ and $Y$ fail in sequence.\\

Since the probability is defined for sets, we define a \texttt{DFT\_event} that satisfies the condition that the input function is less than or equal to time $t$, which represents the moment of time until which we are interested in finding the probability of failure.

\begin{mdframed}
\begin{flushleft}
\begin{definition}
\label{DEF:DFT_event}
\emph{}\\
{\small
\textup{\texttt{$\vdash$ $\forall$p X t. DFT\_event p X t = \{s | X s $\leq$ Normal t\} $\cap$ p\_space p}}}
\end{definition}
\end{flushleft}
\end{mdframed}

\noindent where \texttt{Normal} typecasts the type of $t$ from \texttt{real} to \texttt{extreal}, $p$ represents the probability space and $X$ represents the time-to-failure function.\\

We formally verify the equivalence between the probability of the \texttt{DFT\_event} of an extended real function and its equivalent CDF of the real version of the function~as:

\begin{mdframed}
	\begin{flushleft}
		\begin{theorem}
			\label{THM:CDF_DFT_event}
\emph{ } \\{\small
				\textup{\texttt{$\vdash$ $\forall$X p t. ($\forall$s. X s $\neq$ PosInf $\wedge$ 0 $\leq$ X s) $\Rightarrow$}\\
{\texttt{~~~~(CDF p ($\lambda$s. real (X s)) t = prob p (DFT\_event p X t))}}}}
		\end{theorem}
	\end{flushleft}
\end{mdframed}

\noindent where \texttt{real} is mirror opposite to the typecasting \texttt{Normal} operator. This typecasting is required as the \texttt{DFT\_event} is defined for \texttt{extreal} data-type, and the CDF is defined for real random variables only. Therefore, it is required to ensure that the input function does not equal $+\infty$ and is greater than or equal to 0 since it represents the time of failure of a system component.   

\subsubsection{Probabilistic Behavior of AND, HSP, OR and FDEP Gates}

\vspace{5pt}
To formally verify Equations (\ref{eq:pr_AND}) and (\ref{eq:pr_OR}), we verify the equivalence of the DFT event of the AND gate to the intersection of two events and the OR as the union:

\vspace{5pt}
\begin{mdframed}
	\begin{flushleft}
		\begin{lemma}
			\label{AND_INTER}
\emph{} \\{\small
				\textup{\texttt{$\vdash$ $\forall$p t X Y.}\\
{\texttt{~~~~DFT\_event p (D\_AND X Y) t = DFT\_event p X t $\cap$ DFT\_event p Y t}}}}
		\end{lemma}
	\end{flushleft}
\end{mdframed}

\vspace{5pt}
\begin{mdframed}
	\begin{flushleft}
		\begin{lemma}
			\label{OR_UNION}
\emph{} \\{\small
				\textup{\texttt{$\vdash$ $\forall$p t X Y.}\\
{\texttt{~~~~DFT\_event p (D\_OR X Y) t = DFT\_event p X t $\cup$ DFT\_event p Y t}}}}
		\end{lemma}
	\end{flushleft}
\end{mdframed}

\vspace{70pt}

Based on the independence of random variables and using Theorem \ref{THM:CDF_DFT_event}, we formally verify Equation (\ref{eq:pr_AND}) in HOL4 as:

\vspace{5pt}
\begin{mdframed}
	\begin{flushleft}
		\begin{theorem}
			\label{AND_prob}
\emph{} \\{\small
				\textup{\texttt{$\vdash$ $\forall$p t X Y. rv\_gt0\_ninfty [X; Y] $\wedge$}\\
\mbox{\texttt{~~~~indep\_var p lborel ($\lambda$s. real (X s)) lborel ($\lambda$s. real (Y s)) $\Rightarrow$}}\\
{\texttt{~~~~(prob p (DFT\_event p (D\_AND X Y) t) =}}\\
{\texttt{~~~~~CDF p ($\lambda$s. real (X s)) t * CDF p ($\lambda$s. real (Y s)) t }}}}
		\end{theorem}
	\end{flushleft}
\end{mdframed}

\vspace{5pt}
\noindent where \texttt{indep\_var} ensures the independence of the random variables, $X$ and $Y$, over the Lebesgue-Borel (\texttt{lborel}) measure \cite{Qasim-thesis}. In general, for any two random variables $X$ and $Y$, \texttt{indep\_var} ensures that the probability of the intersection of their events is equal to the multiplication of the probability of the individual events.  Independence of random variables is defined as \cite{Qasim-thesis}:

\begin{mdframed}
	\begin{flushleft}
		\begin{definition}
			\label{indep_vars}
\emph{} \\{\small
				\textup{\texttt{$\vdash$ indep\_vars p M X ii = }\\
{\texttt{~~~($\forall$i. i $\in$ ii $\Rightarrow$}}\\
{\texttt{~~~~~~random\_variable (X i) p}}\\
{\texttt{~~~~~~~~(m\_space (M i), measurable\_sets (M i))) $\wedge$}}\\
{\texttt{~~~indep\_sets p }}\\
{\texttt{~~~~~~($\lambda$i. \{PREIMAGE f A $\cap$ p\_space p |}}\\
{\texttt{~~~~~~~~~(f = X i) $\wedge$  A $\in$ measurable\_sets (M i)\}) ii}}}}
		\end{definition}
	\end{flushleft}
\end{mdframed}

\noindent where \texttt{p} is the probability space, \texttt{M} is the measure space that the random variable~\texttt{X} maps to. In this case, \texttt{M} and \texttt{X} are indexed by a number from the set of numbers~\texttt{ii}, which gives the possibility of defining the independence for multiple random variables that map from the probability space to different spaces. \texttt{indep\_vars} defines the independence by first ensuring that the group of input functions \texttt{X} are random variables and that their event sets are independent using \texttt{indep\_sets}. Using \texttt{indep\_sets}, the probability of the intersection of any sub-group of events of the random variables is equal to the multiplication of the probability of the individual events. 

Using \texttt{indep\_vars}, the independence of two random variables is defined as \cite{Qasim-thesis}:

\begin{mdframed}
	\begin{flushleft}
		\begin{definition}
			\label{indep_var}
\emph{} \\{\small
				\textup{\texttt{$\vdash$ indep\_var p M\_x X M\_y Y = }\\
{\texttt{~~indep\_vars p ($\lambda$i. if i = 0 then M\_x else M\_y) }}\\
{\texttt{~~~~~~~~~~~~~~~($\lambda$i. if i = 0 then X else Y) \{x | (x = 0) $\vee$ (x = 1)\}}}}}

		\end{definition}
	\end{flushleft}
\end{mdframed}

In Theorem \ref{AND_prob}, \texttt{rv\_gt0\_ninfty} adds the condition that the inputs are greater than or equal to 0 but not equal to +$\infty$. We define this generally for the elements of any list as:

\begin{mdframed}
	\begin{flushleft}
		\begin{definition}
			\label{rv_gt0_ninfty}
\emph{} \\{\small
				\textup{\texttt{$\vdash$ {\texttt{(rv\_gt0\_ninfty [] = T) $\wedge$}}\\
{\texttt{~~(rv\_gt0\_ninfty (h::t) = ($\forall$s. 0 $\leq$ h s $\wedge$ h s $\neq$ PosInf) $\wedge$}}\\
{\texttt{~~~~~(rv\_gt0\_ninfty t))}}}}}

		\end{definition}
	\end{flushleft}
\end{mdframed}

This condition is required since we are dealing with the real versions of the random variables. It is a logical condition, since any real-world component will eventually fail, so we are interested only in dealing with the time of failure that is not $\infty$.

\vspace{5pt}
In Theorem \ref{AND_prob}, the random variables are type-casted as real-valued, using the operator \texttt{real}, to function over the Lebesgue-Borel (\texttt{lborel}) measure. \texttt{lborel} is purposely used here to facilitate the Lebesgue integration over the real line when expressing the probabilities of the \textit{before} and \textit{after} events. 

\vspace{5pt}
Theorem \ref{AND_prob} represents the probability of the AND gate and the $HSP$ gate, since based on Theorem \ref{THM:HSP_AND} the behavior of the $HSP$ is equivalent to the behavior of the AND gate. 

\vspace{5pt}
We formally verify Equation (\ref{eq:pr_OR}) based on the probabilistic PIE and  the independence of random variables and using Theorem \ref{THM:CDF_DFT_event} as:

\begin{mdframed}
	\begin{flushleft}
		\begin{theorem}
			\label{OR_prob}
\emph{} \\{\small
				\textup{\texttt{$\vdash$ $\forall$p t X Y. rv\_gt0\_ninfty [X; Y] $\wedge$ All\_distinct\_events p [X;Y] t  $\wedge$}\\
{\texttt{~~~~indep\_var p lborel ($\lambda$s. real (X s)) lborel ($\lambda$s. real (Y s)) $\Rightarrow$ }}\\
{\texttt{~~~~(prob p (DFT\_event p (D\_OR X Y) t) = }}\\
{\texttt{~~~~~CDF p ($\lambda$s. real (X s)) t + CDF p ($\lambda$s. real (X s)) t - }}\\
{\texttt{~~~~~CDF p ($\lambda$s. real (X s)) t $\times$  CDF p ($\lambda$s. real (Y s)) t)}  }}}
		\end{theorem}
	\end{flushleft}
\end{mdframed}

\noindent where \texttt{All\_distinct\_events} ascertains that the event sets are not equal. We formally define it as:
\begin{mdframed}
	\begin{flushleft}
		\begin{definition}
			\label{ALL_DISTINCT_sets2}
\emph{} \\{\small
				\textup{\texttt{$\vdash$ All\_distinct\_events p L t = }\\
{\texttt{~~ALL\_DISTINCT (MAP ($\lambda$x. DFT\_event p x t) L}}}}
		\end{definition}
	\end{flushleft}
\end{mdframed}

\noindent where \texttt{ALL\_DISTINCT} is a HOL4 predicate, which  ensures that the elements of its input list are not equal, \texttt{MAP} is a function that applies the input function \texttt{($\lambda$x. DFT\_event p x t)} to all the elements in the list \texttt{L} and returns a list. This condition is required for the probabilistic PIE.

Theorem \ref{OR_prob} provides the probability of the OR gate as well as the FDEP gate, since the behavior of the FDEP is equivalent to the OR gate according to Theorem~\ref{THM:FDEP2}.

\subsubsection{Probabilistic Behavior of PAND Gate and Before Operator}

We verify Equations (\ref{eq:pr_after}) and (\ref{eq:pr_before}) as Theorems \ref{prob_X_less_Y} and \ref{prob_X_BEFORE_Y}, respectively.

\begin{mdframed}
	\begin{flushleft}
		\begin{theorem}
			\label{prob_X_less_Y}
\emph{}\\			
{\small
				\textup{\texttt{$\vdash$ $\forall$X Y p fy t.}\\
{\texttt{~~~~rv\_gt0\_ninfty [X; Y] $\wedge$ 0 $\leq$ t $\wedge$
     prob\_space p $\wedge$}\\
{\texttt{~~~~indep\_var p lborel ($\lambda$s. real (X s)) lborel ($\lambda$s. real (Y s)) $\wedge$}}\\
{\texttt{~~~~distributed p lborel ($\lambda$s. real (Y s)) fy $\wedge$ ($\forall$y. 0 $\leq$ fy y) $\wedge$}}\\
{\texttt{~~~~cont\_CDF p ($\lambda$s. real (X s)) $\wedge$}}\\
{\texttt{~~~~measurable\_CDF p ($\lambda$s. real (X s)) $\Rightarrow$}}\\
{\texttt{~~~~(prob p (DFT\_event p (Y$\cdot$(X$\lhd$Y)) t) =}}\\
{\texttt{~~~~~pos\_fn\_integral lborel}}\\
{\texttt{~~~~~~~~($\lambda$y. fy y *}}\\
{\texttt{~~~~~~~~~~~~~(indicator\_fn \{w | 0 $\leq$ w $\wedge$ w $\leq$ t\} y *}}\\
{\texttt{~~~~~~~~~~~~~~CDF p ($\lambda$s. real (X  s)) y)))}}}}}
		\end{theorem}
	\end{flushleft}
\end{mdframed}

\begin{mdframed}
	\begin{flushleft}
		\begin{theorem}
			\label{prob_X_BEFORE_Y}
\emph{}\\
{\small
				\textup{\texttt{$\vdash$ $\forall$X Y p fy t.}\\
{\texttt{~~~~rv\_gt0\_ninfty [X; Y] $\wedge$ 0 $\leq$ t $\wedge$ prob\_space p  $\wedge$}\\
{\texttt{~~~~indep\_var p lborel ($\lambda$s. real (X s)) lborel ($\lambda$s. real (Y s)) $\wedge$}}\\
{\texttt{~~~~distributed p lborel ($\lambda$s. real (X s)) fx $\wedge$ ($\forall$x. 0 $\leq$ fx x) $\wedge$ }}\\
{\texttt{~~~~measurable\_CDF p ($\lambda$ s. real (Y s)) $\Rightarrow$}}\\
{\texttt{~~~~(prob p (DFT\_event p (X $\lhd$ Y) t) =}}\\
{\texttt{~~~~~pos\_fn\_integral lborel}}\\
{\texttt{~~~~~~~~($\lambda$x. fx x * }}\\
{\texttt{~~~~~~~~~~~~~~(indicator\_fn \{u | 0 $\leq$ u $\wedge$ u $\leq$ t\} x *}}\\
{\texttt{~~~~~~~~~~~~~~~(1- CDF p ($\lambda$s real (Y  s)) x)))}}}}}		
        \end{theorem}
	\end{flushleft}
\end{mdframed}

\begin{sloppypar}

\noindent where \texttt{pos\_fn\_integral} is the Lebesgue integral for positive functions \cite{Mhamdi-entropy}, \texttt{fy} and \texttt{fx} are the PDF of random variables of the real version of functions $Y$ and $X$, respectively.  \texttt{measurable\_CDF} is a predicate which ensures that the CDF function is a measurable function from the real-borel  space (\texttt{borel}) of the real line to the extreal-borel space (\texttt{Borel}) of the extreal line \texttt{(measurable\_CDF p X = ($\lambda$x. CDF X x) $\in$ measurable borel Borel)}.  \texttt{cont\_CDF} is another predicate, which adds the condition that the CDF is continuous \texttt{(cont\_CDF p X = $\forall$z. ($\lambda$x. real (CDF X x)) contl z)}. This condition is required in Theorem \ref{prob_X_less_Y} as we need to prove that $Pr(X \leq t)$ and $Pr(X < t)$ are equal, and this is not valid unless the CDF function is continuous (\texttt{cont}). 

\end{sloppypar}

Verifying Theorems \ref{prob_X_less_Y} and \ref{prob_X_BEFORE_Y} is not a straightforward task due to the involvement of Lebesgue integration. We  first prove the probability of sets of real random variables in the form of integration before extending the proofs to extended real functions. 

\subsubsection*{Proof Strategy for Theorem \ref{prob_X_less_Y}}

To verify Theorem \ref{prob_X_less_Y}, we first express the event set that corresponds to the integration in Equation (\ref{eq:pr_after}) as:

\begin{equation}
\label{eq:set_after}
(X,Y)^{-1}\{(u,w)\ |\ u\ <\ w\ \wedge\ 0\leq\ w \wedge\ w\ \leq\ t\} 
\end{equation}

Then we verify that the probability of this set can be written using the integration as in Equation (\ref{eq:pr_after}). Therefore, we verify the relationship between the distribution and the integration of positive functions using the push forward measure (\texttt{distr}):

\begin{mdframed}
	\begin{flushleft}
		\begin{theorem}
			\label{distribution_lebesgue_thm2_pos_fn}
\emph{}\\
{\small
				\textup{\texttt{$\vdash$ $\forall$X p M A.}\\
{\texttt{~~~~measure\_space M $\wedge$}\\
{\texttt{~~~~random\_variable X p (m\_space M, measurable\_sets M) $\wedge$}}\\
{\texttt{~~~~A $\in$ measurable\_sets M $\Rightarrow$ }}\\
{\texttt{~~~~(distribution p X A =}}\\
{\texttt{~~~~~pos\_fn\_integral (distr p M X) (indicator\_fn A))}}}}}
        \end{theorem}
	\end{flushleft}
\end{mdframed}

We use Theorem \ref{distribution_lebesgue_thm2_pos_fn} to verify the relationship between the probability and the integration of the joint distribution $F_{XY}$ of two independent random variables as:
\begin{equation}
Pr (X,Y)^{-1}(A) = \int \textbf{1}_{A}\ dF_{XY}
\end{equation}

\indent We formalize this relationship in HOL4 and use a property, which converts the distribution of a pair measure of independent measures into the pair measure of the individual distributions \cite{Qasim-thesis}, to split the integral of joint distributions into two integrals of the individual distributions $(\int \int \textbf{1}_{A}dF_{X} dF_{Y})$. In order to reach the final form of Equation (\ref{eq:pr_after}), we express it in the form of two integrals: 
%&= \int_{A} dF_{X} dF_{Y}\label{eq_int1}\\

\begin{subequations}
\label{eq_eq_int}
\begin{align}
\int^{t}_{0} f_{Y} (y) \times F_{X} (y)\ dy 
&=  \int^{t}_{0} \int_{-\infty}^{y} f_{Y} (y) \times f_{X}(x) \ dx\ dy\label{eq_int2}\\
&= \int^{t}_{0} f_{Y} (y) \ \Big(\int_{-\infty}^{y} f_{X}(x) \ dx\ \Big) dy\label{eq_int3}
\end{align}
\end{subequations}  

\indent The problem in Equations (\ref{eq_int2}) and (\ref{eq_int3}) lies in the fact that the outer integral is a function of the inner integral, i.e., for the inner integral we are integrating until $y$ which is the variable of the outer integral. To be able to handle this formally, we verify that the indicator function of the set in Equation (\ref{eq:set_after}) can be written in the form of the multiplication of two indicator functions, where one is a function of the other. 

\begin{mdframed}
	\begin{flushleft}
		\begin{lemma}
			\label{indicator_of_indicator_after}
\emph{}\\
{\small
				\textup{\texttt{$\vdash$ $\forall$x y t.}\\
{\texttt{~~~~indicator\_fn \{(u,w) | u < w $\wedge$ 0 $\leq$ w $\wedge$ w $\leq$ t\}(x,y)~=}\\
{\texttt{~~~~indicator\_fn \{w| 0 $\leq$ w $\wedge$ w $\leq$ t\} y * indicator\_fn \{u|u < y}\}~x}}}}

        \end{lemma}
	\end{flushleft}
\end{mdframed}

In order to use the above-mentioned lemma and the set on the left hand side, we need to verify that this set is measurable in the two dimensional borel space, i.e., the set belongs to the measurable sets of \texttt{pair\_measure lborel lborel}. This property can be verified based on the fact that the countable union of measurable sets is also measurable. We verify this fact on the rational numbers $\mathbb{Q}$ as follows:

\begin{mdframed}
	\begin{flushleft}
		\begin{theorem}
			\label{measure_space_bigunion_Q}
\emph{}\\
{\small
				\textup{\texttt{$\vdash$ $\forall$m s.}\\
{\texttt{~~~~measure\_space m $\wedge$ ($\forall$n. n $\in$ Q\_set $\Rightarrow$ s n $\in$ measurable\_sets m) $\Rightarrow$}}\\
{\texttt{~~~~BIGUNION (IMAGE s Q\_set) $\in$ measurable\_sets m}}}}
        \end{theorem}
	\end{flushleft}
\end{mdframed}

\noindent where $m$ in our case is \texttt{pair\_measure lborel lborel}. 
The purpose of using the set of rational numbers is that we need a countable set that can be used to express the set in Lemma \ref{indicator_of_indicator_after} as the union of borel rectangles. We verify this in HOL4 as:

\begin{mdframed}
	\begin{flushleft}
		\begin{lemma}
			\label{after_set_bigunion_Q}
\emph{}\\
{\small
				\textup{\texttt{$\vdash$ $\forall$t.}\\
{\texttt{~~~~BIGUNION}}\\
{\texttt{~~~~~~~\{\{u | u < real q\} $\times$ \{w | real q < w $\wedge$ 0 $\leq$ w $\wedge$ w $\leq$ t\} |}}\\
{\texttt{~~~~~~~~~q  $\in$ Q\_set\} = }}\\
{\texttt{~~~~\{(u,w) | u < w $\wedge$ 0 $\leq$ w $\wedge$ w $\leq$ t\}}}}}
        \end{lemma}
	\end{flushleft}
\end{mdframed}

Besides this, we also verify a lemma that the sets in the union of Lemma \ref{after_set_bigunion_Q} are measurable sets in the \texttt{pair\_measure lborel lborel} as:

\begin{mdframed}
	\begin{flushleft}
		\begin{lemma}
			\label{after_set_bigunion_in_measurable_sets}
\emph{}\\
{\small
				\textup{\texttt{$\vdash$ $\forall$t q.}\\
{\texttt{~~~~\{u | u < real q\} $\times$ \{w | real q < w $\wedge$ 0 $\leq$ w $\wedge$ w $\leq$ t\} $\in$}}\\
{\texttt{~~~~measurable\_sets (pair\_measure lborel lborel}}}}
        \end{lemma}
	\end{flushleft}
\end{mdframed}

We can use the proof steps of the previous lemmas to verify the same properties for similar sets, which is essential for other gates expressions. This facilitates dealing with other events in the future, by following the steps in our proof.

By using the above lemmas, we can reason that the original set is measurable set in the \texttt{pair\_measure lborel lborel} as:

\begin{mdframed}
	\begin{flushleft}
		\begin{lemma}
			\label{after_set_in_measurable_sets_pair_lborel}
\emph{}\\
{\small
				\textup{\texttt{$\vdash$ $\forall$t.}\\
{\texttt{~~~~\{(u,w) | u < w $\wedge$ 0 $\leq$ w $\wedge$ w $\leq$ t\} $\in$}}\\
{\texttt{~~~~measurable\_sets (pair\_measure lborel lborel)}}}}
      \end{lemma}
	\end{flushleft}
\end{mdframed}

\vspace{5pt} 

We use Lemmas~\ref{indicator_of_indicator_after} and~\ref{after_set_in_measurable_sets_pair_lborel} to verify that the expression given in Equation~(\ref{eq_int3}) is equal to $\int_{A} dF_{X} dF_{Y}$, where $A$ is the set that specifies the boundaries of the integral. We verify this in HOL4 using the push forward measure as:  
%\begin{equation}
%\label{eq:one_to_two_2}
%\begin{split}
%\int^{t}_{0} f_{Y} (y) \times F_{X} (y)\ dy  &=\int_{A} dF_{X} dF_{Y}\\
%&=\int^{t}_{0} f_{Y} (y) \ \Big(\int_{-\infty}^{y} f_{X}(x) \ dx\ \Big) dy
%\end{split}
%\end{equation}  

\begin{mdframed}[nobreak=true]
	\begin{flushleft}
		\begin{lemma}
			\label{lemma2_indicator_mul_after}
\emph{}\\
{\small
				\textup{\texttt{$\vdash$ $\forall$X Y p t.}\\
{\texttt{~~~~prob\_space p $\wedge$ indep\_var p lborel X lborel Y  $\Rightarrow$}}\\
{\texttt{~~~~(pos\_fn\_integral (pair\_measure (distr p lborel X)}}\\
{\texttt{~~~~~~~~~~~~~(distr p lborel Y))}}\\
{\texttt{~~~~~~~($\lambda$(x,y). indicator\_fn\{(u,w) |u < w $\wedge$ 0 $\leq$ w $\wedge$ w $\leq$ t \}(x,y) = }}\\
{\texttt{~~~~~pos\_fn\_integral (distr p lborel Y)}}\\
{\texttt{~~~~~~~~($\lambda$y. indicator\_fn \{w|0 $\leq$ w $\wedge$ w $\leq$ t\} y *}}\\
{\texttt{~~~~~~~~~~~pos\_fn\_integral(distr p lborel X)}}\\
{\texttt{~~~~~~~~~~~~~~($\lambda$x. indicator\_fn \{u | u < y\} x)))}}}}

        \end{lemma}
	\end{flushleft}
\end{mdframed}
\vspace{5pt}

\noindent where \texttt{indep\_var} ensures that the random variables defined from the probability space $p$ to the borel space are independent, and \texttt{pair\_measure (distr p lborel X) (distr p lborel Y)}, represents the joint distribution of the push forward measures of random variables $X$ and $Y$ over the borel space.

\indent We verify several essential properties for CDF in order to prove that the inner integral of Lemma~\ref{lemma2_indicator_mul_after} is equal to $F_{X}(y)$ or formally to \texttt{(CDF p X y)}. In order to have the PDF of random variable $Y$ in the integral, we assume that the random variable $Y$ has a PDF by definig a density measure for $Y$. We ported the following definition, \texttt{distributed}, from Isabelle/HOL\cite{Johannes-Thesis}, where $f$ in this definition is the PDF of random variable $X$, and the measure part of the density measure is the integral of this PDF. Using this definition, the integral of $f$ is equal to the distribution of the random variable $X$.

\begin{mdframed}[nobreak=true]
	\begin{flushleft}
		\begin{definition}
			\label{distributed}
\emph{}\\
{\small
				\textup{\texttt{$\vdash$ $\forall$p M X f.}\\
{\texttt{~~~~distributed p M X f $\Leftrightarrow$ }\\
{\texttt{~~~~X $\in$}}\\
{\texttt{~~~~measurable(m\_space p,measurable\_sets p)}}\\
{\texttt{~~~~~~(m\_space M,measurable\_sets M) $\wedge$}}\\
{\texttt{~~~~f $\in$ measurable(m\_space M,measurable\_sets M) Borel $\wedge$}}\\
{\texttt{~~~~AE M \{x | 0 $\leq$ f x\} $\wedge$ (distr p M X = density M f)}}}}} 

        \end{definition}
	\end{flushleft}
\end{mdframed}

\begin{sloppypar}
We also use a theorem that replaces the integration with respect to the density measure by the PDF with respect to the original measure (\texttt{lborel} in our case) \cite {Johannes-Thesis}. In addition to the previously verified theorems, we also prove some additional properties, such as sigma finite measure for the push forward measure over the borel space (\texttt{sigma\_finite\_measure (distr p lborel X)}). We also verify that the space generated by the pair measure of two distributions over the borel space is sigma algebra  (\texttt{sigma\_algebra (m\_space (pair\_measure (distr p lborel X)(distr p lborel Y)), measurable\_sets (pair\_measure (distr p lborel X) (distr p lborel Y)))}). Moreover, we verify that the space generated by the space and the measurable sets of the pair measure of \texttt{lborel} is also a sigma algebra (\texttt{{sigma\_algebra (m\_space (pair\_measure lborel lborel), measurable\_sets (pair\_measure lborel lborel))}}). Finally, we prove that the set of the left-hand side of Equation (\ref{eq:pr_after}) is equal to the set that corresponds to the integration of the right-hand side of the same equation as:
\end{sloppypar}

\begin{mdframed}[nobreak=true]
	\begin{flushleft}
		\begin{lemma}
			\label{DFT_event_AND_BEFORE}
\emph{}\\
{\small
				\textup{\texttt{$\vdash$ $\forall$p t X Y.}\\
{\texttt{~~~~rv\_gt0\_ninfty [X; Y] $\wedge$ 0 $\leq$ t $\Rightarrow$}\\
{\texttt{~~~~(DFT\_event p (Y$\cdot$(X$\lhd$Y)) t = }}\\
{\texttt{~~~~~PREIMAGE ($\lambda$x. (real (X x), real (Y x)))}}\\
{\texttt{~~~~~~~~\{(u,w) | u < w $\wedge$ 0 $\leq$ w $\wedge$ w $\leq$ t\} $\cap$ p\_space p}}}}}

        \end{lemma}
	\end{flushleft}
\end{mdframed}

\noindent where \texttt{rv\_gt0\_ninfty} ensures that the input functions are greater than or equal to 0 but not equal to +$\infty$. Based on all the above mentioned lemmas, we are able to verify the original goal for Equation (\ref{eq:pr_after}) as in Theorem \ref{prob_X_less_Y}.\\
%
%PROB_AFTER;
%val it =
%   |- !X Y p fy t.
%     rv_gt0_ninfty X Y /\ 0 <= t /\ prob_space p /\
%     indep_var p lborel (\s. real (X s)) lborel (\s. real (Y s)) /\
%     (!t.
%        {(w,u) | w < u /\ 0 <= u /\ u <= t} IN
%        measurable_sets (pair_measure lborel lborel)) /\
%     distributed p lborel (\s. real (Y s)) fy /\ (!y. 0 <= fy y) /\
%     cont_CDF p (\s. real (X s)) /\
%     measurable_CDF p (\s. real (X s)) ==>
%     (prob p (DFT_event p (D_AND Y (D_BEFORE X Y)) t) =
%      pos_fn_integral lborel
%        (\y.
%           fy y *
%           (indicator_fn {u | 0 <= u /\ u <= t} y *
%            CDF p (\s. real (X s)) y)))

\subsubsection*{Proof Strategy for Theorem \ref{prob_X_BEFORE_Y}}

\indent For the verification of Theorem \ref{prob_X_BEFORE_Y}, we follow almost the same steps for the previous proof. We start by first writing the event set for the integration as: 
\begin{equation}
\label{eq:set_before}
(X,Y)^{-1}\{(u,w)\ |\ 0\ \leq\ u\ \wedge\ u \leq\ t\ \wedge\ u\ <\ w\ \} 
\end{equation}

\indent Then, we describe the indicator function of this set as the multiplication of two indicator functions as: 

\begin{mdframed}[nobreak=true]
	\begin{flushleft}
		\begin{lemma}
			\label{indicator_of_indicator_before}
\emph{}\\
{\small
				\textup{\texttt{$\vdash$ $\forall$x y t.}\\
{\texttt{~~~~indicator\_fn \{(u,w) | 0 $\leq$ u $\wedge$ u $\leq$ t $\wedge$ u < w\}(x,y) = }\\
{\texttt{~~~~indicator\_fn \{u | 0 $\leq$ u $\wedge$ u $\leq$ t\} x * indicator\_fn \{w | x < w\} y}}}}}

        \end{lemma}
	\end{flushleft}
\end{mdframed}

Similar to the procedure, explained previously for the set of the after event in Lemmas \ref{after_set_bigunion_Q}, \ref{after_set_bigunion_in_measurable_sets} and \ref{after_set_in_measurable_sets_pair_lborel}, we verify that the set of the before event is a measurable set in the \texttt{pair\_measure lborel lborel}.

Finally, we rewrite Equation (\ref{eq:pr_before}) as:

\begin{equation}
\label{in_before}
\begin{split}
Pr\{X\lhd Y\}(t) &= \int_{0}^{t} \int_{x}^{\infty} f_{X}(x)\ f_{Y}(y)\  dy\  dx\\
&= \int^{t}_{0} f_{X} (x) \ \Big(\int_{x}^{\infty} f_{Y}(y) \ dy\ \Big) dx
\end{split}
\end{equation}

We verify some additional properties for the CDF in order to complete the proof. For example, we verify that $\int_{x}^{\infty}f_{Y}(y)\ dy$ is equal to $1-F_{Y}(x)$. 
Similarly, we also formally verify that the event of the left-hand side of Equation (\ref{eq:pr_before}) is equal to the set that corresponds to the integration of the right-hand side of the same equation. We use the set in Equation~(\ref{eq:set_before}) to verify this as:

%
%
%val BEFORE_PREIMAGE_event_GT0 =
%   |- !X Y p t.
%     rv_gt0_ninfty X Y /\ 0 <= t ==>
%     (DFT_event p (D_BEFORE X Y) t =
%      PREIMAGE (\s. (real (X s),real (Y s)))
%        {(w,u) | 0 <= w /\ w < u /\ w <= t} INTER p_space p):

\begin{mdframed}[nobreak=true]
	\begin{flushleft}
		\begin{lemma}
			\label{DFT_event_BEFORE}
\emph{}\\
{\small
				\textup{\texttt{$\vdash$ $\forall$p t X Y.}\\
{\texttt{~~~~rv\_gt0\_ninfty [X; Y] $\wedge$ 0 $\leq$ t $\Rightarrow$}\\
{\texttt{~~~~(DFT\_event p (X$\lhd$Y) t = }}\\
{\texttt{~~~~~PREIMAGE ($\lambda$s. (real (X s),real (Y s)))}}\\
{\texttt{~~~~~~~~\{(u,w) | 0 $\leq$ u $\wedge$ u < w $\wedge$ u $\leq$ t\} $\cap$ p\_space p}}}}}
        \end{lemma}
	\end{flushleft}
\end{mdframed}

\noindent Based on all these verified theorems, we are able to formally verify Theorem \ref{prob_X_BEFORE_Y}.
% PROB_DFT_BEFORE;
%val it =
%   |- !X Y p fx t.
%     rv_gt0_ninfty X Y /\ 0 <= t /\ prob_space p /\
%     indep_var p lborel (\s. real (X s)) lborel (\s. real (Y s)) /\
%     distributed p lborel (\s. real (X s)) fx /\ (!x. 0 <= fx x) /\
%     {(w,u) | 0 <= w /\ w <= t /\ w < u} IN
%     measurable_sets (pair_measure lborel lborel) /\
%     measurable_CDF p (\s. real (Y s)) ==>
%     (prob p (DFT_event p (D_BEFORE X Y) t) =
%      pos_fn_integral lborel
%        (\x.
%           fx x *
%           (indicator_fn {w | 0 <= w /\ w <= t} x *
%            (1 - CDF p (\s. real (Y s)) x)))):
%   thm

\vspace{5pt}
So far, we presented the formal verification of the probabilistic behavior of:
\begin{itemize}
\item The AND and HSP gates using Theorem \ref{AND_prob} (since they are equivalent).
\item The OR and FDEP gates using Theorem \ref{OR_prob} (since they are equivalent). 
\item The PAND gate for basic events using Theorem \ref{prob_X_less_Y}. 
\item The \emph{Before} operator using Theorem \ref{prob_X_BEFORE_Y}. 
\end{itemize}
There is no probability of failure for the \textit{Simultaneous} operator as it is eliminated for basic events according to Equation (\ref{simult_never}). This implies that the probability  of the \textit{Inclusive Before} operator is equal to the probability of the \textit{Before} operator for basic events. 

\subsection{Probabilistic Behavior of Gates with Dependent Events}
\label{Pr_dependent_events}

The probabilistic behavior of the CSP and WSP requires dealing with dependent events, as the failure of the main part affects the behavior of the spare part. Therefore, it is required to approach the proof in a different manner.\\

For the $CSP$, the failure distribution of the spare part is affected by the failure time of the main part, as the cold spare starts working after the failure of the main part. Hence, the failure distribution of the spare part is dependent on the failure of the main part. The probability of failure for the output event of a CSP with $Y$ as the main part and $X$ as the spare part is given by \cite {Merle-thesis}:

\begin{equation}
\label{eq:CSP_prob}
Pr(Q_{CSP})(t) = \int_{0}^{t} \Big{(}\int_{v}^{t} f_{(X_{a}|Y=v)} (u) du \Big{)} f_{Y}(v) dv
\end{equation}

\noindent where $f_{(X_{a}|Y=v)} $ is the conditional probability density function for the spare part in its active state ($X_{a}$) given that the main part($Y$) has failed at time $v$. It can be noticed from Equation (\ref{eq:CSP_prob}) that the failure distribution of the spare part is affected by the failure of the main part. Hence, these two input events are not independent, and we cannot utilize the previously verified relationships in Section \ref{Pr_independent_events} to verify the probabilistic behavior of the CSP gate. \\

For the WSP gate with two basic events, the output fails in two cases, Case 1: when the main part fails, then the spare fails in its active state (this case is similar to the CSP case); Case 2: when the spare part fails in its dormant state, then the main part fails with no spare to replace it. In the latter case, the failure distribution of the spare part in its dormant state is independent of the main part. Hence, we can use the previously verified expressions for this case. The probability expression for a WSP with $X$ as the spare part ($X_{a}$ for the active state and $X_{d}$ for the dormant state) and $Y$ as the main part is expressed as \cite{Merle-thesis}:

\begin{equation}
\label{eq:WSP_prob}
Pr(Q_{WSP})(t) = \int_{0}^{t} \Big{(}\int_{v}^{t} f_{(X_{a}|Y=v)} (u) du \Big{)} f_{Y}(v) dv + \int_{0}^{t} f_{Y}(u) F_{X_{d}}(u) du
\end{equation}

\noindent where $F_{X_{d}}$ is the CDF of $X$ in its dormant state.  The first part of Equation (\ref{eq:WSP_prob}) represents the probability of a CSP and the second part represents the probability when the spare fails before the main part. For the second part, $Y$ and $X_{d}$ are independent and hence we can use Equation (\ref{eq:pr_after}) for this case.\\

We verify Equations (\ref{eq:CSP_prob}) and (\ref{eq:WSP_prob}) as Theorems \ref{thm:CSP_prob} and \ref{thm:WSP_prob}, respectively. 

\begin{mdframed}[nobreak=true]
	\begin{flushleft}
		\begin{theorem}

			\label{thm:CSP_prob}
\emph{}\\
{\small
				\textup{\texttt{$\vdash$ $\forall$p X Y f\_xy f\_y f\_cond t.}\\
{\texttt{~~~~rv\_gt0\_ninfty [X; Y] $\wedge$ 0 $\leq$ t $\wedge$}}\\		
{\texttt{~~~~($\forall$y. cond\_density lborel lborel p }}\\
{\texttt{~~~~~~~~~~($\lambda$s. real (X s)) ($\lambda$s. real (Y s)) y f\_xy f\_y f\_cond) $\wedge$}}\\
{\texttt{~~~~prob\_space p $\wedge$ den\_gt0\_ninfty f\_xy f\_y f\_cond  $\Rightarrow$}}\\
{\texttt{~~~~(prob p (DFT\_event p (CSP Y X) t) = }}\\
{\texttt{~~~~~pos\_fn\_integral lborel}}\\
{\texttt{~~~~~~~~($\lambda$y.}}\\
{\texttt{~~~~~~~~~~~indicator\_fn \{u | 0 $\leq$ u $\wedge$ u $\leq$ t\} y * f\_y y * }}\\
{\texttt{~~~~~~~~~~~pos\_fn\_integral lborel}}\\
{\texttt{~~~~~~~~~~~~~~($\lambda$x. indicator\_fn \{w | y < w $\wedge$ w $\leq$ t\} x * f\_cond y x )))}}}}
  				
        \end{theorem}
	\end{flushleft}
\end{mdframed}

\begin{mdframed}[nobreak=true]
	\begin{flushleft}
		\begin{theorem}

			\label{thm:WSP_prob}
\emph{}\\
{\small
				\textup{\texttt{$\vdash$ $\forall$p Y X\_a X\_d t f\_y f\_xy f\_cond.}\\
{\texttt{~~~~prob\_space p $\wedge$ ($\forall$s. ALL\_DISTINCT [X\_a s; X\_d s; Y s]) $\wedge$ }}\\
{\texttt{~~~~(D\_AND X\_a X\_d = NEVER) $\wedge$ rv\_gt0\_ninfty [X\_a; X\_d; Y] $\wedge$ 0 $\leq$ t $\wedge$ }}\\
{\texttt{~~~~($\forall$y. cond\_density lborel lborel p }}\\
{\texttt{~~~~~~~~~~($\lambda$s. real (X\_a s))($\lambda$s. real (Y s)) y f\_xy f\_y f\_cond) $\wedge$}}\\
{\texttt{~~~~den\_gt0\_infty f\_xy f\_y f\_cond $\wedge$}}\\
{\texttt{~~~~indep\_var p lborel ($\lambda$s. real (X\_d s)) lborel ($\lambda$s. real (Y s)) $\wedge$}}\\
{\texttt{~~~~cont\_CDF p ($\lambda$s. real (X\_d s)) $\wedge$}}\\
{\texttt{~~~~measurable\_CDF p ($\lambda$s. real (X\_d s)) $\Rightarrow$}}\\
{\texttt{~~~~(prob p (DFT\_event p (WSP Y X\_a X\_d) t) = }}\\
{\texttt{~~~~~pos\_fn\_integral lborel}}\\
{\texttt{~~~~~~~~($\lambda$y.}}\\
{\texttt{~~~~~~~~~~~indicator\_fn \{u | 0 $\leq$ u $\wedge$ u $\leq$ t\} y * f\_y y * }}\\
{\texttt{~~~~~~~~~~~pos\_fn\_integral lborel}}\\
{\texttt{~~~~~~~~~~~~~~($\lambda$x. indicator\_fn \{w | y < w $\wedge$ w $\leq$ t\} x * f\_cond y x ))+ }}\\
{\texttt{~~~~~pos\_fn\_integral lborel }}\\
{\texttt{~~~~~~~~($\lambda$y.}}\\
{\texttt{~~~~~~~~~~~f\_y y * }}\\
{\texttt{~~~~~~~~~~~(indicator\_fn \{u | 0 $\leq$ u $\wedge$ u $\leq$ t\} y * }}\\
{\texttt{~~~~~~~~~~~~CDF p ($\lambda$s. real (X\_d s)) y )))}}}}
  				
        \end{theorem}
	\end{flushleft}
\end{mdframed}

\noindent where $p$ is the probability space, \texttt{f\_xy} is the joint density function for $X$ and $Y$, \texttt{f\_y} is the marginal density function for $Y$, \texttt{cond\_density} defines the conditional density function (\texttt{f\_cond}) for~$X$ given that ($Y = y$) and \texttt{den\_gt0\_ninfty} ensures the proper values for the density functions and is defined as:

\begin{mdframed}[nobreak=true]
	\begin{flushleft}
		\begin{definition}
			\label{def:den_gt0}
\emph{}\\
{\small
				\textup{\texttt{$\vdash$ $\forall$f\_xy f\_y f\_cond.}\\
{\texttt{~~~~den\_gt0\_ninfty f\_xy f\_y f\_cond $\Leftrightarrow$}}\\
{\texttt{~~~~$\forall$x y. }}\\
{\texttt{~~~~~~0 $\leq$ f\_xy (x,y) $\wedge$ 0 < f\_y y $\wedge$ f\_y y $\neq$ PosInf $\wedge$ 0 $\leq$ f\_cond y x}}}}
        \end{definition}
	\end{flushleft}
\end{mdframed}

\vspace{5pt}
It is noticed that the spare part in the CSP is used without any subscript, i.e., it is used as $X$, since the spare has only one state in the CSP, which is the active state. Therefore, there is no need to use any subscript to distinguish between the dormant and the active states. While in the WSP, we need to distinguish between the two states, i.e., active and dormant, hence the usage of $X_{a}$ and $X_{d}$. For Theorem \ref{thm:WSP_prob}, the condition \texttt{D\_AND X\_a X\_d = NEVER} ensures that the spare part can only fail in one of its states but not both. In addition, it is assumed that the spare part in the dormant ($X_{d}$) state is independent of the main part $Y$ since the failure of the spare part in its dormant state is not affected by the failure of the main part.  As with the previous theorems in Section \ref{Pr_independent_events}, we need to use the typecast operator \texttt{real} with the random variables, since the random variables are of type \texttt{extreal} and the integral over the \texttt{lborel} requires real random variables. 

\subsubsection*{Proof Strategy for Theorem \ref{thm:CSP_prob} (CSP Gate)}

In order to verify Theorem \ref{thm:CSP_prob}, we formalize the conditional density function as \cite{probability-theory-book}:

\begin{mdframed}[nobreak=true]
	\begin{flushleft}
		\begin{definition}

			\label{def:cond_density}
\emph{}\\
{\small
				\textup{\texttt{$\vdash$ $\forall$M1 M2 p X Y y f\_xy f\_y f\_cond.}\\
{\texttt{~~~~cond\_density M1 M2 p X Y y f\_xy f\_y f\_cond $\Leftrightarrow$}}\\
{\texttt{~~~~random\_variable X p (m\_space M1, measurable\_sets M1) $\wedge$}}\\
{\texttt{~~~~random\_variable Y p (m\_space M2, measurable\_sets M2) $\wedge$}}\\
{\texttt{~~~~distributed p (pair\_measure M1 M2) ($\lambda$x. (X x, Y x)) f\_xy $\wedge$}}\\
{\texttt{~~~~distributed p M2 Y f\_y $\wedge$ (f\_cond y = ($\lambda$x. f(x,y) / f\_y y)) 
}}}}
  				
        \end{definition}
	\end{flushleft}
\end{mdframed}

\noindent where \texttt{p} is the probability space, \texttt{M1} and \texttt{M2} are the measure spaces that the random variables $X$ and $Y$ map to, respectively (we will use \texttt{lborel} in our case), \texttt{f\_xy} is the joint density function for $X$ and $Y$, \texttt{f\_y} is the marginal density function of $Y$ and finally, \texttt{f\_cond} is the conditional density function of $X$ given ($Y = y$).
 
The conditional density function definition ensures that $X$ and $Y$ are random variables with joint density function \texttt{f\_xy} and a marginal density function \texttt{f\_y}. It is noticed from the definition of the conditional density function \texttt{f\_cond} that it is a function of $x$ only, and it can have different variants depending on the value of $Y$ that we are conditioning at, i.e., $y$. This is why \texttt{f\_cond} takes $y$ as a parameter.   

From Definition \ref{def:cond_density}, we formally verify the following relationship between the conditional density, the joint density and the marginal density functions, given that $f_{Y}(y) \neq 0$:

\begin{equation}
f_{XY} (x,y) = f_{X|Y=y} (x) \times f_{Y}(y)
\end{equation}

The above equation can be formalised in HOL4 as:

\begin{mdframed}[nobreak=true]
	\begin{flushleft}
		\begin{theorem}

			\label{thm:cond_joint_marginal_density}

\emph{}\\
{\small
\textup{\texttt{$\vdash$ $\forall$M1 M2 p X Y f\_cond x y f\_xy f\_y. }\\
{\texttt{~~~~($\forall$y. f\_y y $\neq$ 0 $\wedge$ f\_y y $\neq$ PosInf $\wedge$ f\_y y $\neq$ NegInf) $\wedge$}}\\
{\texttt{~~~~cond\_density M1 M2 p X Y y f\_xy f\_y f\_cond $\Rightarrow$}}\\
{\texttt{~~~~(f\_xy (x,y) = f\_cond y x * f\_y y)
}}}}
 				
        \end{theorem}
	\end{flushleft}
\end{mdframed}

The condition \texttt{f\_y y $\neq$ 0} is required, as this function will be used in the denominator of the conditional density and it cannot equal to 0. In addition, since we are dealing with extended-real numbers, \texttt{f\_y y} cannot equal infinity.

The second step in verifying the expression of the CSP is by verifying that the probability of the joint random variables is equal to the iterated integrals of the joint density function. This can be expressed as:

\begin{equation}
Pr(X,Y)^{-1}(A) = \int\int \textbf{1}_{A} \times f_{XY}(x,y)dx\ dy
\end{equation}

We use Theorem \ref{distribution_lebesgue_thm2_pos_fn} to verify this in HOL4 as:

\begin{mdframed}[nobreak=true]
	\begin{flushleft}
		\begin{theorem}

			\label{thm:prob_joint_density_iterated_integrals}

\emph{}\\
{\small
\textup{\texttt{$\vdash$ $\forall$p X Y f\_xy A. }\\
{\texttt{~~~~distributed p (pair\_measure lborel lborel) ($\lambda$x. (X x, Y x)) f\_xy $\wedge$}}\\
{\texttt{~~~~prob\_space p $\wedge$ ($\forall$x. 0 $\leq$ f\_xy x) $\wedge$}}\\
{\texttt{~~~~A $\in$ measurable\_sets (pair\_measure lborel lborel)$\Rightarrow$}}\\
{\texttt{~~~~(prob p (PREIMAGE ($\lambda$x. (X x, Y x)) A $\cap$ p\_space p) = }}\\
{\texttt{~~~~~pos\_fn\_integral lborel}}\\
{\texttt{~~~~~~~~($\lambda$y.}}\\
{\texttt{~~~~~~~~~~pos\_fn\_integral lborel}}\\
{\texttt{~~~~~~~~~~~~~($\lambda$x. indicator\_fn A (x,y) * f\_xy (x,y))))}}}}
		
        \end{theorem}
	\end{flushleft}
\end{mdframed}

Then, we express the probability of the joint random variables using the conditional density function as:

\begin{equation}
Pr(X,Y)^{-1}(A) = \int\int \textbf{1}_{A} \times f_{(X|Y=y)}(x) \times f_{Y} (y)\ dx\ dy
\end{equation}

We verify this in HOL4, using Theorems \ref{thm:cond_joint_marginal_density} and \ref{thm:prob_joint_density_iterated_integrals}, as:

\begin{mdframed}[nobreak=true]
	\begin{flushleft}
		\begin{theorem}

			\label{thm:prob_cond_density_iterated_integrals}

\emph{}\\
{\small
\textup{\texttt{$\vdash$ $\forall$p X Y f\_xy  f\_y f\_cond A. }\\
{\texttt{~~~~($\forall$y. cond\_density lborel lborel p X Y y f\_xy f\_y f\_cond) $\wedge$}}\\
{\texttt{~~~~prob\_space p $\wedge$ ($\forall$x. 0 $\leq$ f\_xy x) $\wedge$}}\\
{\texttt{~~~~($\forall$y. 0 < f\_y y $\wedge$ f\_y y $\neq$ PosInf) $\wedge$}}\\
{\texttt{~~~~A $\in$ measurable\_sets (pair\_measure lborel lborel)$\Rightarrow$}}\\
{\texttt{~~~~(prob p (PREIMAGE ($\lambda$x. (X x, Y x)) A $\cap$ p\_space p) = }}\\
{\texttt{~~~~~pos\_fn\_integral lborel}}\\
{\texttt{~~~~~~~~($\lambda$y.}}\\
{\texttt{~~~~~~~~~~pos\_fn\_integral lborel}}\\
{\texttt{~~~~~~~~~~~~~($\lambda$x. indicator\_fn A (x,y) * f\_cond y x * f\_ y y )))}}}}
		
        \end{theorem}
	\end{flushleft}
\end{mdframed}

In order to be able to reach the final form of Equation (\ref{eq:CSP_prob}), we need first to express the event set that corresponds to the integration in Equation (\ref{eq:CSP_prob}) as:

\begin{equation}
(X,Y)^{-1}\{(x,y)\ |\ y\ <\ x \wedge\ x\ \leq\ t\ \wedge\ 0\ \leq\ y\ \wedge\ y\ \leq\ t\}
\end{equation}

We verify in HOL4 that this set corresponds to the \texttt{DFT\_event} of the CSP gate~as:

\begin{mdframed}[nobreak=true]
	\begin{flushleft}
		\begin{lemma}

			\label{lem:CSP_PREIMAGE_event}

\emph{}\\
{\small
\textup{\texttt{$\vdash$ $\forall$X Y p t. }\\
{\texttt{~~~~rv\_gt0\_ninfty [X; Y] $\wedge$ 0 $\leq$ t} $\Rightarrow$}\\
{\texttt{~~~~(DFT\_event p (CSP Y X) t = }}\\
{\texttt{~~~~~PREIMAGE ($\lambda$s. (real (X s), real (Y s)))}}\\
{\texttt{~~~~~~~~\{(x,y)| y < x $\wedge$ x $\leq$ t $\wedge$ 0 $\leq$ y $\wedge$ y $\leq$ t\} $\cap$ p\_space p)}}}}
		
        \end{lemma}
	\end{flushleft}
\end{mdframed}

 \vspace{-1pt}
In addition, we verify that the event set in Lemma \ref{lem:CSP_PREIMAGE_event} is measurable in \texttt{pair\_measure lborel lborel}.
Finally, we verify that the indicator function of the set in Lemma \ref{lem:CSP_PREIMAGE_event} can be expressed as the multiplication of two indicator functions to determine the boundaries of the iterated integrals in Equation (\ref{eq:CSP_prob}) as:

\begin{mdframed}[nobreak=true]
	\begin{flushleft}
		\begin{lemma}

			\label{lem:CSP_indicator_of_indicator}

\emph{}\\
{\small
\textup{\texttt{$\vdash$ $\forall$x y t. }\\
{\texttt{~~~~indicator\_fn \{(w,u) | u < w $\wedge$ w $\leq$ t $\wedge$ 0 $\leq$ u $\wedge$ u $\leq$ t\} (x,y) = }}\\
{\texttt{~~~~indicator\_fn \{w | y < w $\wedge$ w $\leq$ t\} x *}}\\
{\texttt{~~~~indicator\_fn \{u | 0 $\leq$ u $\wedge$ u $\leq$ t\} y}}}}
		
        \end{lemma}
	\end{flushleft}
\end{mdframed}
 
 \vspace{-1pt}
Using all these verified theorems and lemmas, we formally verify Theorem \ref{thm:CSP_prob}.
\vspace{-3pt}
\subsubsection*{Proof Strategy for Theorem \ref{thm:WSP_prob} (WSP Gate)}

For the verification of Theorem \ref{thm:WSP_prob}, it is evident that the probability expression involves the probability of the CSP gate in addition to the probability of the \textit{after} expression of Theorem \ref{prob_X_less_Y}. Therefore, we choose to verify that the event of the WSP for basic events is equivalent to the union of two sets as:

\begin{mdframed}[nobreak=true]
	\begin{flushleft}
		\begin{lemma}
			\label{lem:WSP_DFT_event}

\emph{}\\
{\small
\textup{\texttt{$\vdash$ $\forall$p Y X\_a x\_d t. }\\
{\texttt{~~~~($\forall$s. 0 $\leq$ Y s) $\wedge$ ($\forall$s. ALL\_DISTINCT [X\_a s; X\_d s; Y s]) $\wedge$}}\\
{\texttt{~~~~(D\_AND X\_a X\_d = NEVER) $\Rightarrow$}}\\
{\texttt{~~~~(DFT\_event p (WSP Y X\_a X\_d) t = }}\\
{\texttt{~~~~~~\{s | Y s < X\_a s $\wedge$ X\_a s $\leq$ Normal t $\wedge$}}\\ 
{\texttt{~~~~~~~~~~0 $\leq$ Y s $\wedge$ Y s $\leq$ t\} $\cap$ p\_space p $\cup$}}\\
{\texttt{~~~~~~\{s | X\_d s < Y s $\wedge$ Y s $\leq$ Normal t \} $\cap$ p\_space p)}}}}
        \end{lemma}
	\end{flushleft}
\end{mdframed}
 
Then, we verify that the above two sets are disjoint. As these two sets are disjoints then the probability of the original set is equivalent to the sum of the probabilities of the disjoint sets. Based on this, we verify that the probability of the first set (\texttt{\{s | Y s < X\_a s $\wedge$ X\_a s $\leq$ Normal t $\wedge$ 0 $\leq$ Y s $\wedge$ Y s $\leq$ t\} $\cap$ p\_space p}) is equal to the probability of the CSP gate, which will result in the first term in the addition of the conclusion of Theorem \ref{thm:WSP_prob}. We also verify that the probability of the second set in Lemma \ref{lem:WSP_DFT_event}  (\texttt{\{s | X\_d s < Y s $\wedge$ Y s $\leq$ Normal t\} $\cap$ p\_space p)}) is expressed using Theorem \ref{prob_X_less_Y}, which will result in the second term of the addition of the conclusion of Theorem \ref{thm:WSP_prob}. As a result, we have the probability of the WSP as in Theorem \ref{thm:WSP_prob}.

In this section, we formally verified the probabilistic behavior of the DFT gates: AND, OR,  HSP, FDEP, PAND, CSP, WSP and the Before operator  besides the formalization of expressions for $Pr (X < Y \wedge Y \leq t)$ and $Pr (X < Y \wedge X \leq t)$. 

These verified properties are generic, i.e., universally quantified for all distribution and density functions, and can be used to formally verify the probability of failure expression of any DFT. The HOL4 proof script for this verification as well as the gate definitions is available at \cite{ICFEM-code}.    
\section{Formal Verification of the Cardiac Assist System}
\label{Experiment results}
\begin{figure}[!b]
\centering
{\includegraphics[scale=0.4]{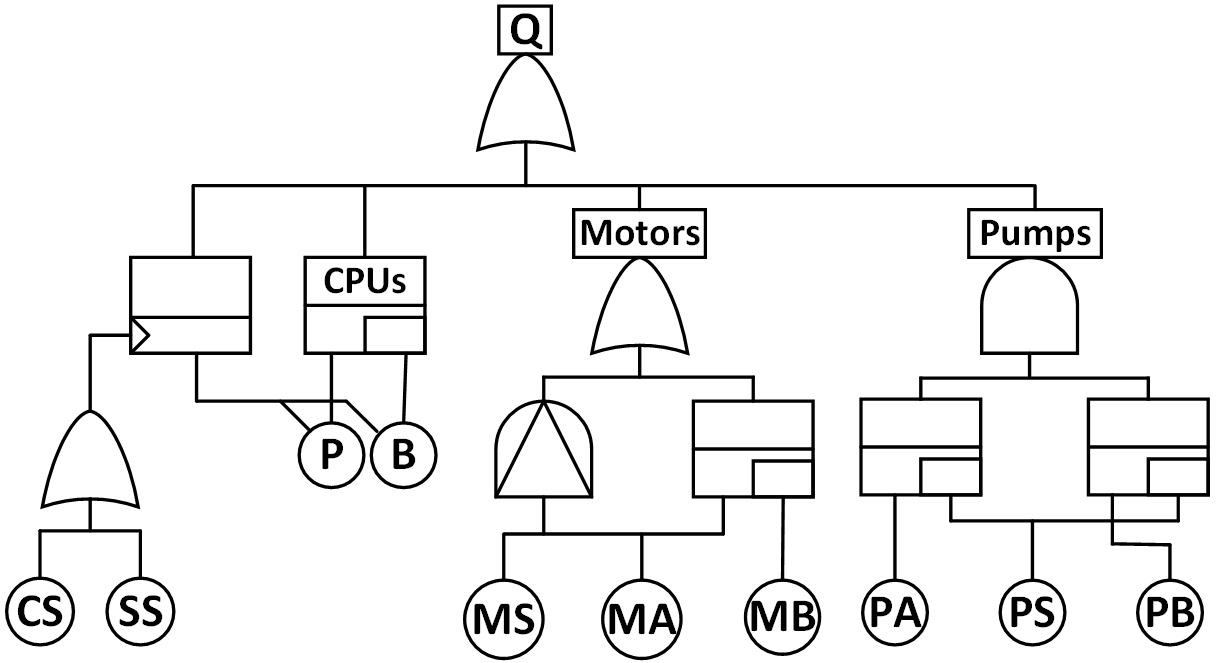}}
\caption{Cardiac Assist System}
%\vspace{-20pt}
\label{fig:CAS}
\end{figure}

In order to illustrate the utilization of our formalized probabilistic behavior of the gates and operators in the last section, we present a DFT-based formal failure analysis of the Cardiac Assist System, shown in Figure \ref{fig:CAS} \cite{hichem-rigorous2010}.

We first provide generic steps that can be followed in order to use our formalization of the DFTs to conduct the formal analysis of DFTs in the form of generic expressions of failure probabilities. These steps are:
\begin{enumerate}

\item Determine the structure function of the top event of the DFT.
\item Simplify the structure function and formally verify that the simplified version is equal to the original function obtained in step (1).
\item Create the \texttt{DFT\_event} of the structure function.
\item Express the \texttt{DFT\_event} of the top event as the union of multiple input events.  
\item \label{apply-PIE} Apply the probabilistic PIE to the union of events generated in the previous step, then simplify the result of the PIE. This will result in having the summation of the probabilities of the intersection of the different events that contribute to the failure of the top event of the DFT.
\item Replace each term in the result of the PIE by its probabilistic expression based on the verified expressions in Section \ref{probabilistic_gates} for each gate and operator. 
\end{enumerate}

Step (\ref{apply-PIE}) requires proving many lemmas that are necessary for manipulating the result of the PIE. For example, we need to verify the associativity property of addition for a large group of numbers (in case of the Cardiac Assist system, we verified this property for 63 numbers). Although this seems a trivial task, it requires dealing with \texttt{extreal} numbers, which includes proving that for all combinations of the inputs, the result of the addition cannot equal to $\infty$. Step~(\ref{apply-PIE}) also requires verifying the power set of events in a recursive way. Moreover, based on the independence of the input random variables, we need to verify the independence of several combinations of random variables (in the Cardiac Assist system, we verified that any two random variables out of the ten are independent, then three out of ten,... etc).

In the rest of this section, we illustrate the utilization of the previous steps to perform the formal DFT analysis of the Cardiac Assist System to provide a generic expression for the probability of failure of the top event.
The Cardiac Assist system consists of three main subsystems: pumps, motors and CPUs. The system has two main pumps ($PA$ and $PB$) with a shared spare $PS$. It has three motors $MS$, $MA$, and $MB$, where $MB$ replaces $MA$ after failure. Finally, the system has one main CPU ($P$) and a spare CPU ($B$). Both CPUs are functionally dependent on a trigger, which is the union of the crossbar switch ($CS$) and the system supervisor ($SS$). In this case study, we are assuming that the spare gates are HSPs. %
%D_OR
%        (D_OR
%           (D_OR (D_OR (D_OR CS SS) (D_AND MA (D_BEFORE MS MA)))
%              (D_AND MA MB)) (D_AND P B)) (D_AND (D_AND PA PB) PS)):
%   thm

Our goal is to verify the probability of of failure of the Cardiac Assist system by applying the probabilistic PIE considering that the input events are independent. This can be represented mathematically as:
\begin{flalign}
\label{eq:prob_CAS}
\begin{split}
Pr(Q) = &F_{CS} (t)\ +\ F_{SS} (t)\ +\ \int_{0}^{t}f_{MA}(y) \times F_{MS} (y)\ dy\ +\\
&F_{MA}(t)\times F_{MB}(t)\ +\ F_{P}(t)\times F_{B}(t)\ +\ F_{PA}(t)\times F_{PB}(t)\times F_{PS}(t)\\
&  -...+...-\ F_{CS} (t)\ \times F_{SS} (t) \times \Big (\int_{0}^{t}f_{MA}(y) \times F_{MS} (y)\ dy\ \Big) \\
& \times F_{MA}(t) \times  F_{MB}(t)\times F_{P}(t)\times F_{B}(t)\times F_{PA}(t)\times F_{PB}(t)\times F_{PS}(t)
\end{split}
\end{flalign}
We verify Equation (\ref{eq:prob_CAS}) for generic probability CDF and PDF in HOL4 as:

%indep_vars_sets [CS;SS;MA;MS;MB;P;B;PA;PB;PS] p t/\
%(!t. {(u,a) | u < a /\ 0<=a/\a <= t} IN
%        measurable_sets (pair_measure lborel lborel)) /\
%     distributed p lborel (\s. real (MA s)) f_MA /\ (!y. 0 <= f_MA y) /\
%     cont_CDF p (\s. real (MS s))/\
% measurable_CDF p (\s. real (MS s))  ==>
%       (prob p (DFT_event p
%       (D_OR
%        (D_OR
%           (D_AND (shared_spare PA PB PS PS)
%              (shared_spare PB PA PS PS))
%           (D_OR (P_AND MS MA) (HSP MA MB)))
%        (HSP (FDEP (D_OR CS SS) P) (FDEP (D_OR CS SS) B)))
%       t)

\begin{mdframed}[nobreak=true]
	\begin{flushleft}
		\begin{theorem}
			\label{PROB_CAS}
\emph{}\\
{\small
				\textup{\texttt{$\vdash$ $\forall$CS SS MA MS MB P B PA PB PS p t f\_MA. }\\
{\texttt{~~~~0 $\leq$ t $\wedge$ prob\_space p $\wedge$}}\\ 
 {\texttt{~~~~ALL\_DISTINCT\_RV [CS;SS;MA;MS;MB;P;B;PA;PB;PS] p t $\wedge$}}\\
 {\texttt{~~~~indep\_vars\_sets [CS;SS;MA;MS;MB;P;B;PA;PB;PS] p t $\wedge$}}\\
{\texttt{~~~~distributed p lborel ($\lambda$s. real (MA s)) f\_MA $\wedge$ ($\forall$y. 0 $\leq$ f\_MA y) $\wedge$}}\\
{\texttt{~~~~cont\_CDF p ($\lambda$s. real (MS s)) $\wedge$}}\\
{\texttt{~~~~measurable\_CDF p ($\lambda$s. real (MS s)) $\Rightarrow$}}\\
{\texttt{~~~~(prob p}}\\
{\texttt{~~~~~~~(DFT\_event p }}\\
{\texttt{~~~~~~~~~~~((shared\_spare PA PB PS PS)$\cdot$(shared\_spare PB PA PS PS)+}}\\
{\texttt{~~~~~~~~~~~(PAND MS MA)+(HSP MA MB)+}}\\
{\texttt{~~~~~~~~~~~(HSP (FDEP(CS + SS) P)(FDEP(CS + SS) B))) t) =}}\\ 
{\texttt{~~~~~CDF p ($\lambda$s. real (CS s)) t + CDF p ($\lambda$s. real (SS s)) t +}}\\
{\texttt{~~~~~pos\_fn\_integral lborel}}\\
{\texttt{~~~~~~~~($\lambda$y.}}\\
{\texttt{~~~~~~~~~~~f\_MA y * (indicator\_fn \{u | 0 $\leq$ u $\wedge$ u $\leq$ t\} y *}}\\
{\texttt{~~~~~~~~~~~~CDF p ($\lambda$s. real (MS s)) y)) +}}\\
{\texttt{~~~~~CDF p ($\lambda$s. real (MA s)) t * CDF p ($\lambda$s. real (MB s)) t +}}\\
{\texttt{~~~~~CDF p ($\lambda$s. real (P s)) t * CDF p ($\lambda$s. real (B s)) t +}}\\
{\texttt{~~~~~CDF p ($\lambda$s. real (PA s)) t * CDF p ($\lambda$s. real (PB s)) t *}}\\
{\texttt{~~~~~CDF p ($\lambda$s. real (PS s)) t - ....+...-}}\\
{\texttt{~~~~~CDF p ($\lambda$s. real (CS s)) t * CDF p ($\lambda$s. real (SS s)) t *}}\\
{\texttt{~~~~~pos\_fn\_integral lborel}}\\
{\texttt{~~~~~~~~($\lambda$y.}}\\
{\texttt{~~~~~~~~~~~f\_MA y * (indicator\_fn \{u | 0 $\leq$ u $\wedge$ u $\leq$ t\} y *}}\\
{\texttt{~~~~~~~~~~~~CDF p ($\lambda$s. real (MS s)) y)) *}}\\
{\texttt{~~~~~CDF p ($\lambda$s. real (MB s)) t * CDF p ($\lambda$s. real (P s)) t *}}\\
{\texttt{~~~~~CDF p ($\lambda$s. real (B s)) t * CDF p ($\lambda$s. real (PA s)) t *}}\\
{\texttt{~~~~~CDF p ($\lambda$s. real (PB s)) t * CDF p ($\lambda$s. real (PS s)) t)}}}}
        \end{theorem}
	\end{flushleft}
\end{mdframed}

\noindent where \texttt{0 $\leq$ t} ensures that the time $t$ is greater than or equal to 0, \texttt{prob\_space p} indicates that $p$ is a probability space, \texttt{ALL\_DISTINCT\_RV} is a predicate which ensures that all inputs and their event sets are not equal and their values are greater than or equal to 0 but they cannot equal +$\infty$. This assumption is a realistic one, since for any component in a system the time of failure will always be greater than or equal to 0 and the component will eventually fail. The predicate \texttt{indep\_vars\_sets} adds the condition that all random variables and their event sets are independent. The predicate (\texttt{distributed p lborel ($\lambda$s. real (MA s)) f\_MA}) indicates that the real random variable of $MA$ has the density function \texttt{f\_MA}. The last two predicates in the goal ensures that the CDF of the real random variable of $MS$ is continuous and measurable from the real line to the extreal one (real-borel to extreal-borel).

\indent We verify several intermediate lemmas to prove Theorem \ref{PROB_CAS}. We first verify a reduced form of the given  DFT and, then we verify the probability expression of the verified reduced version.

\begin{mdframed}[nobreak=true]
	\begin{flushleft}
		\begin{lemma}
			\label{CAS_reduced}
\emph{}\\
{\small
				\textup{\texttt{$\vdash$ $\forall$CS SS MA MS MB P B PA PB PS.}\\
{\texttt{~~~~($\forall$s. ALL\_DISTINCT [MA s; MS s; PA s; PB s; PS s]) 
$\Rightarrow$}}\\
{\texttt{~~~~((shared\_spare PA PB PS PS)$\cdot$(shared\_spare PB PA PS PS) +}}\\
{\texttt{~~~~~(PAND MS MA) +}}\\
{\texttt{~~~~~(HSP MA MB)+(HSP (FDEP(CS + SS) P)(FDEP(CS + SS) B)) =}}\\
{\texttt{~~~~~CS + SS + (MA$\cdot$(MS $\lhd$ MA)) + MA$\cdot$MB + P$\cdot$B + PA$\cdot$PB$\cdot$PS)}}}}
        \end{lemma}
	\end{flushleft}
\end{mdframed}

In the above lemma, \texttt{(shared\_spare PA PB PS PS)$\cdot$(shared\_spare PB PA PS PS)} represents the pumps part of the DFT, \texttt{
(PAND MS MA)+(HSP MA MB)} represents the motors parts and finally the CPUs part is represented by \texttt{(HSP (FDEP(CS + SS)~P)(FDEP(CS + SS) B)}.  
The predicate \texttt{ALL\_DISTINCT} ensures that all basic events cannot fail at the same time. Since we assumed that all spare gates are HSPs, the spare input $PS$ for the shared spare gates is the same for both the active and dormant states.
In order to find the probability of the top event, we utilize the formally verified reduced version of the structure function and encapsulate it into a \texttt{DFT\_event}, as the probability can only be applied to sets. To utilize the probabilistic PIE, we  express the \texttt{DFT\_event} of the Cardiac Assist system as the union of events.

\begin{mdframed}[nobreak=true]
	\begin{flushleft}
		\begin{lemma}
			\label{CAS_union_list}
			
\emph{}\\
{\small
				\textup{\texttt{$\vdash$ $\forall$PA PB PS MS MA MB CS SS P B p t.}\\
{\texttt{~~~~DFT\_event p}}\\
{\texttt{~~~~~~~(CS + SS + (MA$\cdot$(MS $\lhd$ MA)) + MA$\cdot$MB + P$\cdot$B + PA$\cdot$PB$\cdot$PS) t =}}\\
{\texttt{~~~~union\_list}}\\
{\texttt{~~~~~[DFT\_event p CS t; DFT\_event p SS t; }}\\
{\texttt{~~~~~~DFT\_event p (MA$\cdot$(MS $\lhd$ MA)) t;}}\\
{\texttt{~~~~~~DFT\_event p (MA$\cdot$MB) t; }}\\
{\texttt{~~~~~~DFT\_event p (P$\cdot$B) t; DFT\_event p (PA$\cdot$PB$\cdot$PS) t]}}}}
        \end{lemma}
	\end{flushleft}
\end{mdframed}

From Lemma \ref{CAS_union_list}, we can notice that the top event is constructed from the union of six different sets. Applying the probabilistic PIE on the union of these sets (6 sets) generates 63 different terms (combinations). We verify several lemmas to be able to use the theorem of the probabilistic PIE \cite{ahmed2015towards} for the union list of these six sets. For example, we formally verify that

\begin{mdframed}[nobreak=true]
	\begin{flushleft}
		\begin{lemma}
			\label{PIE_set_alt_form_lem_6}
\emph{}\\
{\small
				\textup{\texttt{$\vdash$ $\forall$A B C D E K.}\\
{\texttt{~~~~\{t | t SUBSET \{A; B; C; D; E; k\} $\wedge$ t $\neq$ \{\}\} =}}\\
{\texttt{~~~~\{\{A\}; \{B\}; \{C\}; \{D\}; \{E\}; \{k\}; \{A; B\}; \{A; C\};...;}}\\
{\texttt{~~~~~\{A; B; C; D; E; k\}\}}}}}
        \end{lemma}
	\end{flushleft}
\end{mdframed}

\noindent The result of Lemma \ref{PIE_set_alt_form_lem_6} is a set of 63 different sets. We had to apply the \texttt{SIGMA} function that results from the \texttt{sum\_set} in the PIE theorem. Therefore we verify the following lemma for 63 sets.

\begin{mdframed}[nobreak=true]
	\begin{flushleft}
		\begin{lemma}
			\label{PIE_lem4_63_prod}
\emph{}\\			
{\small
				\textup{\texttt{$\vdash$ $\forall$A B C D E K.}\\ 
{\texttt{~~~~ALL\_DISTINCT [A;B;C;D;E;k] $\wedge$}\\
{\texttt{~~~~($\forall$x. x $\in$\{\{A\};\{B\};\{C\};\{D\};\{E\};\{k\};...;\{A; B; C; D; E; k\}\} $\Rightarrow$}}\\
{\texttt{~~~~~~~f x $\neq$ PosInf) $\Rightarrow$}}\\
{\texttt{~~~~(SIGMA f \{\{A\};\{B\};...;\{A; B; C; D; E; k\}\} =}}\\
{\texttt{~~~~~f \{A\} + f \{B\} +...+ f \{A; B; C; D; E; k\}}}}}}

        \end{lemma}
	\end{flushleft}
\end{mdframed}

%               
% We had also to verify the add associativity property for 63 terms, which was not straight forward as we are dealing with \texttt{extreal} data-type and it required to prove that all the results of the addition of the subsets of these 63 terms cannot equal to +$\infty$ or to -$\infty$.  
After verifying all these lemmas and based on the reduced DFT expression we are able to verify the probability of the Cardiac Assist system (Equation \ref{eq:prob_CAS}) into Theorem \ref{PROB_CAS}.

\begin{sloppypar}
The first part of the conclusion of Theorem~\ref{PROB_CAS} corresponds to the original DFT (without reduction). In the verification of this theorem, we use Lemma~\ref{CAS_reduced} to replace the original DFT with the reduced one. Then, we use Lemma~\ref{CAS_union_list} to represent the \texttt{DFT\_event} as a union list. After representing the left-hand side of the conclusion of Theorem~\ref{PROB_CAS} as a union list, we use Lemmas~\ref{PIE_set_alt_form_lem_6}, \ref{PIE_lem4_63_prod} and the probabilistic PIE theorem \cite{ahmed2015towards} to prove this goal. The first 6 terms in the right-hand side of the conclusion of Theorem~\ref{PROB_CAS} correspond to the probability of the elements of the list in Lemma~\ref{CAS_union_list}. For example, \texttt{CDF p ($\lambda$s. real (CS s)) t} represents the probability of \texttt{DFT\_event~p~CS~t}, which is $F_{CS}(t)$, according to Theorem~\ref{THM:CDF_DFT_event}. \texttt{pos\_fn\_integral lborel($\lambda$y.~f\_MA y~*(indicator\_fn~\{u~|0~$\leq$~u~$\wedge$~u~$\leq$~t\}~y * CDF p ($\lambda$s. real (MS s)) y))} represents the probability of \texttt{DFT\_event~p~(MA$\cdot$(MS~$\lhd$~MA))~t}, which is $\int_{0}^{t} f_{MA}(y)\ \times\ F_{MS}(y)\ dy$, according to Theorem~\ref{prob_X_less_Y}. The following terms in the conclusion of Theorem~\ref{PROB_CAS} correspond to finding the probability of the intersection of each pair in the list, then each 3 elements then 4 elements until we reach the last term in the right-hand side of the goal, which corresponds to the probability of the intersection of all elements in the list. Since all six elements in the union list are independent, the probability of their intersection is equal to the multiplication of the individual probabilities. \\
\indent It is important to note that we have been able to verify the probability of the Cardiac Assist system for generic distributions and density functions, which can be instantiated later with specific functions according to the required constraints, without any need to repeat the whole process from the beginning. It is worth mentioning that such results cannot be obtained using PMCs, as they can only generate the probability of failure after specifying the failure rates of the components. In addition, PMCs are only limited to exponential distribution which does not consider the aging factor in any system. However, using our formalization for generic expression, it can be used with any probability distribution and density function as long as they are integrable, which makes it a more general and accurate alternative to the existing techniques. 

\end{sloppypar}

%\vspace{-20pt}
\section{Conclusions }
\label{Conclusion}

In this work, we proposed to conduct the probabilistic analysis of DFTs within the HOL4 theorem prover and thus obtain formally verified  probability of failure expressions for generic probability distributions and density functions. We verified many simplification theorems for DFT gates and operators that allow formal reasoning about the reduction of the structure function of the DFT top event into a simpler form.  In particular, we verified the probability of the intersection and the union of independent events to provide the probability of the AND, OR, FDEP and HSP gates. Moreover, we verified the probability of a sequence of two failing events $(Pr(X<Y))$ in two forms, i.e, $Pr (X < Y \wedge Y \leq t)$ and $Pr (X < Y \wedge X \leq t)$, which, to the best of our knowledge, is another novel contribution. These expressions are used to formally express the probability of the PAND gate and the before operator. Similarly, we also verified the probabilistic behavior of the spare gates, which required dealing with dependent events and conditional density functions. To illustrate the effectiveness of our formalization, we presented the formal analysis of the Cardiac Assist System, which is a safety-critical system. Using our formalization, we were able to provide generic results for the probability of failure of this system, i.e., for any distributions and density functions. It is evident that such results cannot be obtained using simulation nor using model checking. This highlights the importance of our proposed work, besides the fact that it inherits the sound and expressive nature of HOL theorem proving.  
\balance
\newpage
\bibliography{mabiblio}{}
\bibliographystyle{IEEEtran}
\end{document}